\documentclass[traditabstract]{aa}
\usepackage[varg]{txfonts}

\usepackage{graphicx}
\usepackage{natbib}
\usepackage{amsmath}
\usepackage{amssymb}
\usepackage{url}

\newcommand{\hd}[0]{HD~181068}
\newcommand{\kep}[0]{\textit{Kepler}}
\newcommand{\xmm}[0]{XMM-\textit{Newton}}
\newcommand{\cts}[0]{\mbox{ct\,s$^{-1}$}}
\newcommand{\ergs}[0]{\mbox{erg\,s$^{-1}$}}
\newcommand{\ergcms}[0]{\mbox{erg\,cm$^{-2}$\,s$^{-1}$}}
\newcommand{\rs}[0]{RS~CVn}

\begin{document}

\title{A multiwavelength study of the hierarchical triple \hd:}
\subtitle{A test bed for studying star-planet-interaction?}
\author{S. Czesla \and K.F. Huber \and P.C. Schneider 
   \and J. H. M. M. Schmitt}
\institute{Hamburger Sternwarte, Universit\"at Hamburg, Gojenbergsweg 112, 21029 Hamburg, Germany}
\date{Received ... / Accepted ... }

\abstract{
\hd\ is the only compact, triply eclipsing, hierarchical triple
system containing a giant star known to date. With its central,
highly-active G-type giant orbited by a close pair of
main-sequence dwarfs, the system is ideal to study tidal interactions. 
We carried out a multiwavelength
study to characterize the magnetic activity of the \hd\ system.
To this end, we obtained in- and out-of-eclipse X-ray snapshots with \xmm\
and an optical spectrum, which we analyzed along with the \kep\ light-curve.
The primary giant shows strong
quiescent X-ray emission at a
level of $2\times 10^{31}$~\ergs, an S-index of $0.41\pm0.01$, and marked
white-light flares releasing up to $6\times 10^{38}$~erg in the \kep-band.
During the second X-ray observation, we found a three-times elevated -- yet
decaying -- level of X-ray emission, which might be due to an X-ray flare. The
high level of magnetic activity is
compatible with the previously reported absence of solar-like oscillations in
the giant, whose atmosphere, however, undergoes tidally-induced
oscillations imposed by the changing configuration of the dwarf-binary. We
found that the driving force exciting these oscillations is comparable to the
disturbances produced by a typical hot Jupiter, making the system
a potential test bed to study the effects of tidal interactions also present
in planetary systems.}

\keywords{Stars: activity, Stars: individual: \hd, X-rays: stars, Planet-star
interactions}
\maketitle

\section{Introduction}
While stellar triple systems are not at all rare, the systems
\hd\ and KOI-126 are the only compact triples with mutually
eclipsing components currently known \citep{Raghavan2010, Carter2011,
Derekas2011}.
Of these two, only \hd\ harbors an evolved giant star in its center. Due to the
absence of an orbital equilibrium state, compact triple systems are ideal
laboratories to study tidal interactions, orbital evolution, and to test models
of stellar evolution \citep[e.g.,][]{Fuller2013}.

The central G-type giant of \hd\ is 
orbited by a close pair of main-sequence dwarfs.
While the dwarf-binary\footnotemark[1] completes a full orbit around the giant
in $45$~d, its components revolve around each other every $0.9$~d
\citep[see Table~\ref{tab:orbProp} and][]{Borkovits2013}.
All mutual eclipses have been observed with high accuracy by the \kep\
satellite \citep{Derekas2011}.
Because the effective temperatures of the three system components are identical
to within about $400$~K (see Table~\ref{tab:starProp}), the transit light-curves
resulting from the eclipse of the giant by the dwarf-binary and vice versa do
not differ substantially in depth.
They can, however, be distinguished through their shape, because the primary transit
light-curve, with the binary passing in front of the giant, yields a rounder
profile resulting from limb darkening. The secondary eclipse, during which the
dwarf binary is entirely occulted by the giant, shows a box-like profile
\citep[e.g.,][]{Derekas2011}.

\footnotetext[1]{Following the convention
of \citet{Borkovits2013}, we dub the giant the A-component and the eclipsing
dwarf-binary the B-component; individual constituents of the B-component are
referred to as Ba and Bb.}

The duration of the phase of total eclipse in the AB system  
depends on the respective geometry.
\citet{Borkovits2013} showed
that the eclipse geometry essentially repeats after five orbits
of the AB system\footnotemark[1], i.e., after about $227$~d. The total
duration of every eclipse -- primary or secondary -- amounts to approximately $1.5$
dwarf-binary orbits or $\approx 1.35$~d in every case.

The HIPPARCOS-measured parallax of \hd\ is $4.02\pm0.4$~mas, resulting
in a distance of $249\pm25$~pc.
\citet{Borkovits2013} derived a preliminary age estimate of $300-500$~Ma for the
\hd\ system by comparing the stellar parameters to evolutionary tracks. The
authors caution, however, that the dwarf radii ``appear to be significantly
larger than expected'', and more sophisticated modeling may be necessary.  

The structure of the \hd\ system is reminiscent of that of  
RS~Canum Venaticorum (\rs) systems -- binary systems consisting of a G- or
K-type giant orbited by a late-type main-sequence or subgiant companion
\citep[][]{Dempsey1993, Hall1976}.
\rs\ systems are among the most active stellar systems in the Galaxy as
evidenced, among others, by pronounced and variable emission-line cores in
\ion{Ca}{II}~H and K and H$\alpha$, starspots, photometric variability, and
X-ray emission \citep[e.g.,][]{Linsky1984, Strassmeier1988, Schmitt1990,
Dempsey1993}. In \rs\ systems with periods shorter than about $30$~d,
rotation and orbital motion are typically
synchronized
\citep{Zahn1977, Scharlemann1982, Dempsey1993, Derekas2011}. In \hd,
\citet{Derekas2011} derived a rotational velocity of
\mbox{$v\sin(i)=14$~km\,s$^{-1}$} for the giant, which, combined with its
radius, suggests that the orbital period of the AB system and the giant's
rotation are also synchronized.

In a Fourier
analysis of the \kep\ light-curve,
\citet{Borkovits2013} found substantial photometric variability
at a period comparable to the AB system's orbital period. The observed
amplitudes exceed the expectation for 
ellipsoidal variation. Because the giant's rotation is likely synchronized with
the AB system's orbital motion, this strongly suggests rotational modulation
due to starspots, which might even be seen as distortions in some
primary-eclipse light-curves \citep[][Fig.~5]{Borkovits2013}.
The authors proceed to argue that their Fourier analysis reveals a double-peak
close to the presumed rotation period of the giant, which may be a consequence of
differential rotation.

Furthermore, the \kep\ light-curve shows a highly interesting behavior at higher
frequencies. First, it reveals a number of oscillation
periods that can be attributed to 
tidal interactions between the changing configuration of the dwarf-binary and
the atmosphere of the giant \citep{Derekas2011, Borkovits2013, Fuller2013}. Such
tidally-induced oscillations can only be observed in sufficiently close systems
such as \hd\ or, potentially, planetary systems. Second, solar-like
oscillations of the central giant are absent or
strongly suppressed \citep{Derekas2011, Fuller2013}. \citet{Chaplin2011} argued
that solar-like oscillations are damped by magnetic activity
in solar-like stars; according to these authors, this certainly seems also to be
the case in the Sun. Therefore, \citet{Fuller2013} speculate that strong
activity may also suppress solar-like oscillations in the central giant \hd~A.
Some support for this argument comes from a number
of pronounced flare-like events observed in the \kep\ light-curve, one of
which is observed during a secondary eclipse
\citep[][Fig.~6]{Borkovits2013} and, thus, must be attributed to the giant, if
it originates in the \hd\ system.
Another indication for a high level of magnetic activity has been contributed by
the ROSAT all-sky survey, which revealed a clear X-ray source with a count
rate of $0.294\pm0.023$~\cts\ at the position of \hd, suggesting an X-ray
luminosity on the order of $2\times 10^{31}$~\ergs\ (cf.,
Sect.~\ref{sec:ROSAT}).

In this paper, we present a multiwavelength study of stellar activity in the
\hd\ system. In particular, we provide a detailed analysis of the X-ray
emission from the \hd\ system based on two carefully timed \xmm\ observations,
study the white-light flares observed in the \kep\ light-curves, and
investigate the chromospheric \ion{Ca}{II} H and K emission using an optical
spectrum obtained during secondary eclipse.

\begin{table}[h]
  \caption{Stellar properties of the system components\protect\footnotemark[1]
  \citep{Borkovits2013}.
  \label{tab:starProp}}
  \begin{tabular}{l l l l}
  \hline\hline
   & A & Ba & Bb \\
   \hline
   m [M$_{\odot}$] & $3\pm0.1$ & $0.915\pm0.034$ & $0.870\pm0.043$  \\
   R [R$_{\odot}$] & $12.46\pm0.15$ & $0.865\pm0.01$ & $0.8\pm0.02$  \\
   T$_{eff}$ [K] & $5100\pm100$ & $5100\pm100$ & $4675\pm100$  \\
   L$_{bol}$ [L$_{\odot}$] & $92.8\pm7.6$ & $0.447\pm0.037$ & $0.27\pm0.027$ \\
   $\log(g)$ [cgs] & $2.73$ & $4.53$ & $4.58$ \\ \hline
  \end{tabular}
\end{table}

\begin{table}[h]
  \caption{Orbital parameters of the \hd\ system\protect\footnotemark[1]
  \citep{Borkovits2013}.
  Numbers in parentheses indicate error in last digits.
  \label{tab:orbProp}}
  \begin{tabular}{l l l}
  \hline\hline
      & A$-$B & Ba$-$Bb \\
  \hline
   T$_{MIN,p}^a$ (BJD$_{TDB}$) & $2\,455\,499.9970(4)$ &
   $2\,455\,051.237\,00(5)$
   \\
   P [d] & $45.4711(2)$ & $0.9056768(2)$ \\
   a [R$_{\odot}$] & $90.31\pm0.72$ & $4.777\pm0.039$ \\
   i [$^{\circ}$] & $87.5\pm0.2$ & $86.7\pm1.4$ \\ \hline
  \end{tabular}
  \tablefoot{$^a$\; Reference time for primary eclipse. Times converted from UTC
  as used by \kep\ and stated in \citet{Borkovits2013} into Barycentric
  Dynamical Time (TDB) by adding $66.184$~seconds (T. Borkovits, private
  communication; see also ``Kepler data release 19 notes'' Sect. 3.4).}
\end{table}

\begin{table}[h]
\caption{Duration (DUR.) and sum of ``good time intervals'' (ONTIME) for the
individual instruments of our \xmm\ observations.
\label{tab:Filtering}}
  \begin{tabular}{l l l l l}
  \hline\hline
  & \multicolumn{2}{c}{0722330201} & \multicolumn{2}{c}{0722330301} \\
  & DUR.$^a$ & ONTIME & DUR.$^a$ & ONTIME \\ 
  Instr. & [ks] & [ks] & [ks] & [ks] \\ 
  \hline
  pn & 13.0 & 11.7 & 12.0 &  8.2\\
  MOS~1 & 14.6 & 14.5 & 13.6 & 13.6\\
  MOS~2 & 14.6 & 14.6 & 13.6 & 13.6\\
  \hline
  \end{tabular}
  \tablefoot{$^a$\;Difference between the \texttt{TSTOP} and
  \texttt{TSTART} times.}
\end{table}

\section{X-ray observations and data analysis}

We observed \hd\ twice using \xmm. While the first X-ray observation was carried
out when both giant and dwarf-binary component were visible, the second
observation had been scheduled during secondary eclipse, when the dwarf-binary
remained hidden behind the giant. The timing details are given in
Table~\ref{tab:xmmObs}, where the phase is defined as a value between
zero and one according to the ephemeres given in Table~\ref{tab:orbProp}.
In both observations, we used
the medium filter for pn and MOS~1 and the thick filter for MOS~2. 

\begin{table}
  \caption{Timing of the \xmm\ observations.
  \label{tab:xmmObs}}
  \begin{tabular}{l l l l}
  \hline\hline
  & UTC$^a$ & JD$_{UTC}^b$ & BJD$_{TDB}^b$ \\
  \multicolumn{4}{c}{Observation ID 0722330201}\\
  \hline
  Start & 2013-05-03 4:26 & $415.684\,72$ & $415.685\,98$\\
  End   & 2013-05-03 8:03 & $415.835\,42$ & $415.836\,68$\\
  Phase & \multicolumn{3}{c}{$0.1378 - 0.1411^c$} \\
  \hline
  \multicolumn{4}{c}{Observation ID 0722330301}\\
  \hline
  Start & 2013-05-19 14:56 & $432.115\,28$ & $432.117\,24$ \\
  End   & 2013-05-19 18:17 & $432.261\,81$ & $432.263\,78$ \\
  Phase & \multicolumn{3}{c}{$0.4992-0.5024^c$} \\ \hline
  \end{tabular}
  \tablefoot{$^a$\; \texttt{DATE-OBS} and \texttt{DATE-END} header keywords for
  the pn; $^b$\;~$2456000$ has been subtracted; $^c$\; The typical error in
  phase is $8.9\times 10^{-5}.$}
\end{table}

We reduced the data using \xmm's ``Scientific Analysis System'' (SAS) in
version 12.0.1 applying standard recipes and filtering
recommendations\footnote{Please
see http://xmm.esac.esa.int/sas/current/documentation/threads/ for the
standard recipes.}. The duration of the observations and the available
time remaining after the filtering are given in
Table~\ref{tab:Filtering}. While the pn-exposure during the in-eclipse phase (ID
0722330301) suffers from high background toward the end of the observation,
the MOS instruments remain virtually unaffected. The X-ray data analysis was
carried out using XSPEC in version 12.5.0 \citep{Arnaud1996}.

\begin{figure}[h]
  \includegraphics[width=0.49\textwidth]{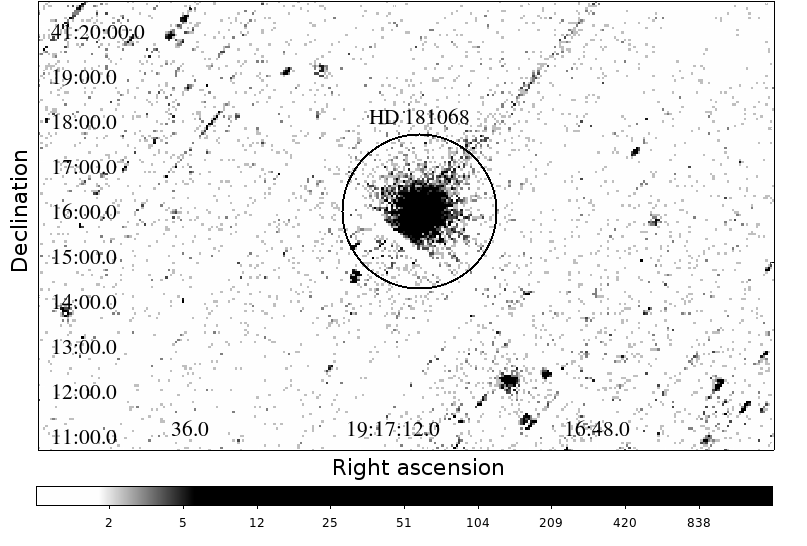}
  \caption{In-eclipse X-ray image of \hd\ (pn).
  \label{fig:PNImage}}
\end{figure}

An X-ray source at the position of \hd\ is evident in both the in- and
out-of-eclipse X-ray images.
Figure~\ref{fig:PNImage} shows the in-eclipse X-ray image of \hd\ observed
by the pn-camera.
In this phase, we even found a
stronger source with a count rate approximately tripled compared to the previous
out-of-eclipse exposure (Table~\ref{tab:PNMOSfit}).

\subsection{Short-term X-ray variability}

\begin{figure}[h]
  \includegraphics[angle=-90,width=0.49\textwidth]{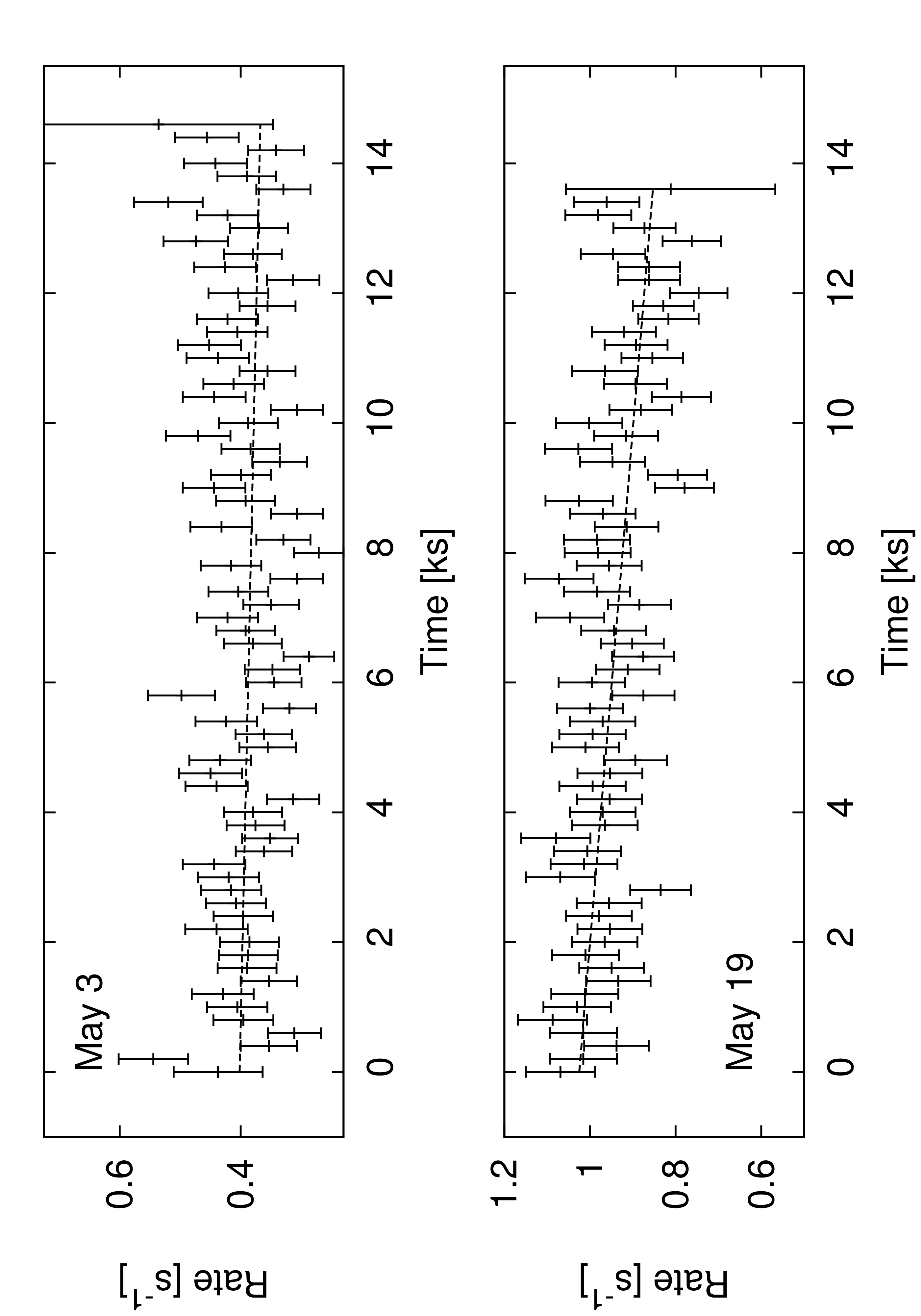}
  \caption{Background-subtracted MOS~1 light-curve of \hd\ in the $0.3-6$~keV
  band with $200$~s binning.
  \label{fig:moslc}}
\end{figure}

To check for short-term X-ray variability in \hd, we constructed $0.3-6$~keV
light-curves using the \texttt{evselect} and \texttt{epiclccorr} routines. The
latter corrects for various effects such as vignetting, bad pixels, good-time
intervals, and dead time. As
an example, we show in Fig.~\ref{fig:moslc} the MOS~1 light-curve
observed during May 3 and 19.

First, we fitted the MOS~1, MOS~2, and pn light-curves using a constant model.
The resulting best-fit count-rates are listed in Table~\ref{tab:linfit}.
Second, we fitted a linear model. While the constant could be varied
independently for the three instruments, the
gradient was taken to be identical for all simultaneous light
curves. In Table~\ref{tab:linfit} the $\chi^2$-values and number of degrees
of freedom (dof) for the best-fit constant and linear model are listed. 
Applying an F-test, we found that the model including the linear term
provides a better description of the data at the $95$\% confidence level in both
cases. In particular, we find p-values of $0.02$ and $10^{-16}$ for May
3 and 19.

The lower level and relative stability of the X-ray count-rate
on May 3 suggest to identify the X-ray flux observed then with the quiescent
X-ray emission level; this interpretation is also consistent with the
ROSAT observation, which yields a similar flux (cf., Sect.~\ref{sec:ROSAT}).
On May 19, we did not only find a higher X-ray luminosity but also a
more strongly declining X-ray count-rate, which seems less typical for
quiescent emission.

During both X-ray observations, \xmm's Optical Monitor (OM) observed
\hd\ with the UVM2 filter in imaging mode. Comparing the images, we derived a
decrease of $0.36\pm0.27$\% in the OM count-rate during the eclipse. This is
roughly compatible with an expected $0.8$\%, assuming that the luminosities of
the system components in the UVM2 filter are proportional to their bolometric
luminosities. The OM provides no evidence for an increased near ultra-violet
count-rate accompanying the elevated X-ray count-rate during the in-eclipse
observation.

\begin{table}[h]
\caption{Fit results for the constant and linear model along with the best-fit
$\chi^2$-values and the number of degrees of freedom (dof).
\label{tab:linfit}}
\begin{tabular}{l l l}
\hline\hline
Instrument & CR$^a$ May 3 & CR$^a$ May 19 \\
           & [ct\,$s^{-1}$] & [ct\,$s^{-1}$] \\ \hline
           & \multicolumn{2}{c}{Best-fit constants} \\
pn   & $1.45\pm 0.013$  & $3.374\pm 0.025$\\ 
MOS~1 & $0.387\pm 0.006$ & $0.938\pm 0.009$ \\
MOS~2 & $0.325\pm 0.005$ & $0.828\pm 0.009$\\ \hline
 &  \multicolumn{2}{c}{Best-fit gradient [ct\,s$^{-1}$\,ks$^{-1}$]} \\
Combined &  $-0.0023\pm0.0009$ & $-0.0125\pm0.0015$ \\ \hline
\multicolumn{3}{c}{$\chi^2$-values and dof} \\
Constant & 239.1/207 & 232.9/206 \\
Linear   & 200.8/179 & 136.8/178 \\ \hline
\end{tabular}
\tablefoot{$^a$\, count rate (CR)}
\end{table}

\subsection{Spectral analysis of pn and MOS data}
\label{sec:PNMOS}

\begin{figure}[h]
  \includegraphics[angle=-90, width=0.49\textwidth]{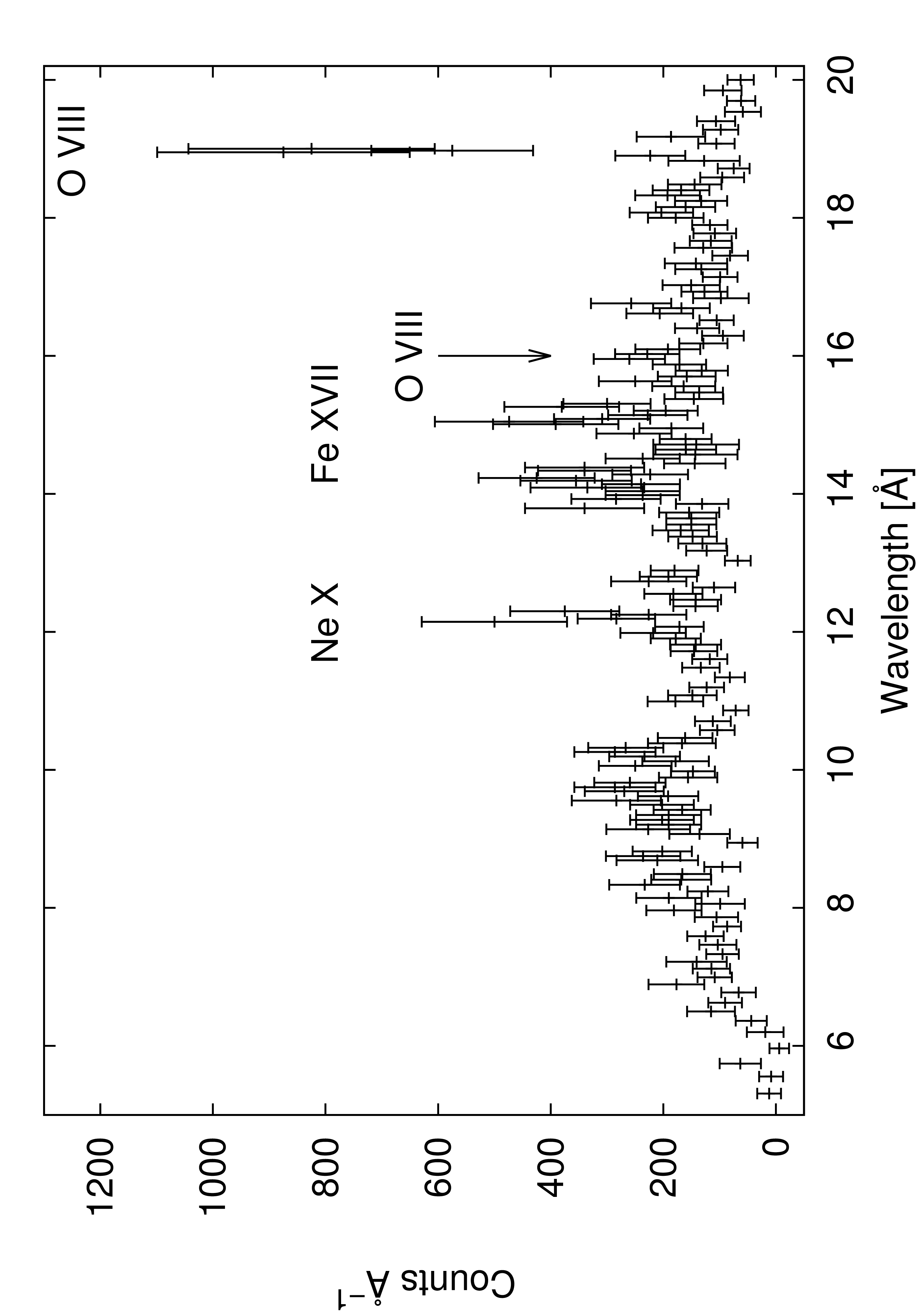}
  \caption{Merged, in-eclipse RGS~1+2 spectrum of \hd\ with 15~counts per
  spectral bin. Labels denote the most prominent spectral lines.
  \label{fig:RGS1_301}}
\end{figure}

The pn count-rate of \hd\ is sufficiently high to produce a source with
substantial signal exceeding the background level even in the wings of the
point-spread-function (PSF). Therefore, we opted for a $30''$-radius source
region to take full advantage of the high flux in our analysis.
Strong X-ray sources are susceptible to pile-up, due to the limited detector
read-out cadence.
The ``\xmm\ users' Handbook'' (Sect. 3.3.2, Table~3) states
critical limits of $8$~\cts\ for the pn and $0.7$~\cts\ for the MOS instruments
operated in ``full frame'' mode, beyond which pile-up starts to seriously
deteriorate the X-ray spectra.
According to these limits, pile-up is no issue for the pn data;
the MOS data may be mildly affected (cf.,
Table~\ref{tab:linfit}). An additional analysis
of the pn-data based on the SAS-tool \texttt{epaplot} also yielded a negligible
pile-up fraction. As we further see
no difference between the MOS data observed with the thick and medium filter,
we consider pile-up and optical loading irrelevant in our analysis.

\begin{table}
  \caption{Best-fit parameters based on pn and MOS data and the RGS data
  with $90$\% confidence intervals.
  \label{tab:PNMOSfit}}
  \begin{tabular}{l l l}
  \hline\hline
  \multicolumn{3}{c}{pn and MOS} \\
      & May 3 & May 19 \\ \hline
  \hline
  N$_H$ [$10^{22}$~cm$^{-2}$] & $0.021_{-0.004}^{+0.004}$ &
  $0.044_{-0.004}^{+0.004}$ \\
  T$_1$ [keV]                 & $0.92_{-0.04}^{+0.04}$ &
  $0.88_{-0.05}^{+0.07}$ \\
  EM$_1^a$ [$10^{53}$~cm$^{-3}$] & $7.5_{-0.3}^{+0.3}$  &
  $7.7_{-0.5}^{+0.5}$ \\
  T$_2$ [keV]                  & $1.88_{-0.11}^{+0.12}$ &
  $1.95_{-0.24}^{+0.48}$ \\
  EM$_2^a$ [$10^{53}$~cm$^{-3}$] & $10.4_{-0.4}^{+0.4} $ &
  $28.6_{-1.3}^{+1.3}$ \\
  T$_3$ [keV]                  &  &
  $8.6_{-2.2}^{+37}$ \\
  EM$_3^a$ [$10^{53}$~cm$^{-3}$] & &
  $12.8_{-0.8}^{+0.8}$ \\
  Ab.$_{Fe,Mg, Al, Ni, Ca}$ [$\odot$] &
  \multicolumn{2}{c}{$0.23_{-0.03}^{+0.02}$} \\
  Ab.$_{C,N,O,S,Si}$ [$\odot$] &
  \multicolumn{2}{c}{$0.16_{-0.05}^{+0.05}$}\\
  Ab.$_{Ne,Ar}$ [$\odot$] &
  \multicolumn{2}{c}{$0.73_{-0.21}^{+0.21}$} \\
  $\log_{10}(\mbox{Flux [cgs]})^{a,b}$ & $-11.5896_{-0.008}^{+0.008} $ & 
  $-11.0877_{-0.006}^{+0.006}$\\ \hline
  $\chi^2$/dof & \multicolumn{2}{c}{$1729.5/1631$} \\
  \hline \hline
  \multicolumn{3}{c}{RGS} \\ \hline
  Ab.$_{Fe,Mg, Al, Ni, Ca}$ [$\odot$] &
  \multicolumn{2}{c}{$0.35_{-0.09}^{+0.11}$} \\
  Ab.$_{C,N,O,S,Si}$ [$\odot$] &
  \multicolumn{2}{c}{$0.55_{-0.14}^{+0.19}$}\\
  Ab.$_{Ne,Ar}$ [$\odot$] &
  \multicolumn{2}{c}{$1.49_{-0.38}^{+0.49}$} \\ \hline
  \end{tabular}
  \tablefoot{$^a$\; The distance uncertainty is not included. $^b$\;
  unabsorbed flux, $0.3-9.0$~keV band}
\end{table}

In our spectral
analysis, the spectra were modeled with an absorbed
thermal model with variable abundances \citep[\texttt{vapec},][]{Smith2001}.
In particular, we used two temperature components for the May 3 observation and
included an additional hot component for the observation on May 19.
This model is motivated by the results of
more detailed grating observations of the \rs\ systems II~Peg and AR~Lac
\citep{Huenemoerder2001, Huenemoerder2003}. While these systems show
complex differential emission measure (DEM) distributions, the
DEM reconstructions also show distinct peaks. In the case of AR~Lac the DEM
peaks at about $0.1$~keV, $0.69$~keV, and $1.9$~keV. In both
cases, the low-temperature end of the DEM remains mostly constant in time,
while variability manifests at the high-temperature end of the DEM. In the
particular case of II~Peg, a flare was observed, which could be modeled as an
addition of a hot component to the DEM without affecting the cooler components
\citep{Huenemoerder2001}.

We treated both observations and the MOS and pn instruments
simultaneously in our spectral analysis.
The pn and MOS spectra were grouped into bins comprising $15$~counts each,
which renders the $\chi^2$-statistic applicable.
In our treatment of the elemental abundances,
we followed
the approach of \citet{Schmitt2007} and grouped the elements with respect to
their first ionization potential (FIP).
In particular, low-FIP (Fe,
Mg, Al, Ni, Ca; FIP$ < 8$~eV), medium-FIP (C, N, O, S, Si; $8$~eV $<$ FIP $<
15$~eV), and high-FIP elements (Ne, Ar; FIP $> 15$~eV) were distinguished; solar
abundances refer to those given by \citet{Anders1989}.

While the abundance pattern was coupled among all thermal components in the fit, 
their temperatures and emission measures could be varied independently. The sum
of the thermal components was subject to absorption represented by the
\texttt{phabs} model, and a different depth of the absorption column was allowed
for the two observations.
To compute the unabsorbed flux and its error, we applied the \texttt{cflux}
model. Please note that the values and errors of the emission measures were
calculated with the \texttt{cflux} component removed and all other parameters
fixed to ensure consistency; the errors may, therefore, be slightly
underestimated.
With this approach we obtained a model with a reduced $\chi^2$ value of $1.06$;
our results are summarized in
Table~\ref{tab:PNMOSfit}.

While the lowest-temperature component characterized by $T_1$ and EM$_1$
remained virtually unchanged between May 3 and May 19, the medium-temperature
component approximately tripled its emission measure, EM$_2$, but remained at
about the same temperature. For the temperature of the third, hot component,
$T_3$, we found a best-fit value of $100$~MK, which remained loosely determined,
however.
Demanding a common absorption column depth during both observations
resulted in a model with reduced $\chi^2$ value of $1.12$, a column depth of
\mbox{$3.8\times 10^{22}$~cm$^{-2}$}, and otherwise similar parameters.
Formally, this solution provides an inferior model, which may indicate the
presence of circumstellar material. However, this interpretation is not unique,
because the corona of
the giant also changed and the geometrical configuration during the two
observations was different. 
The dwarf components are probably also highly active and provided an
unknown and not explicitly accounted for coronal contribution only to the
out-of-eclipse observation.

The fluxes given in Table~\ref{tab:PNMOSfit} correspond to average X-ray
luminosities of $(1.9\pm0.4)\times 10^{31}$~\ergs\ on May 3 and
$(6.1\pm1.2)\times 10^{31}$~\ergs\ on May 19.

As we found evolution in the X-ray count-rate of \hd\ during the observation on
May 19, we also searched for temporal variability in the spectrum. In a
first step, we examined a hardness ratio. In particular,
we defined a low-energy band, $L$, ranging from $0.3-1$~keV and a high-energy
band, $H$, from $1$~keV to $6$~keV, and the associated hardness ratio
$HR=(H-L)/(H+L)$.
Based on the MOS data, we found a mean hardness ratio
of $0.62$. Applying a linear fit, we detected a
decrease of $10^{-2}$~ks$^{-1}$ in hardness ratio, which is significant on the
$99$\% confidence level according to an F-test and is consistent with a
softening of the X-ray spectrum.
The pn data are also consistent with a
decrease in hardness, but do not provide a significant result alone,
because they do not offer the same temporal coverage. Further, the hardness
ratio cannot be directly compared, because the
spectral sensitivity of the pn is different. In a second step, we checked
whether the decrease in hardness can directly be detected in a spectral
analysis.
To this end, we divided the observation into three $4.5$~ks chunks and
repeated the previously described spectral analysis using the events pertaining
to the individual observing chunks.
We re-fitted the spectral model allowing, however, only the emission
measure of the medium- and high-temperature components, EM$_2$ and EM$_3$, to
vary. The remaining parameters remained fixed at the values reported in
Table~\ref{tab:PNMOSfit}. We note that the temperature, $T_3$, of the
high-temperature component was not well defined in any of the chunks. Applying a
linear fit, we found an average decrease of \mbox{$-1.9\times
10^{52}$~cm$^{-3}$\,ks$^{-1}$} for EM$_2$ and \mbox{$-4.1\times
10^{52}$~cm$^{-3}$\,ks$^{-1}$} for EM$_3$, which is compatible with cooling of
the high-temperature plasma component.

\subsection{RGS data}
In our analysis of the RGS data, we used the same spectral model as for the pn
and MOS data (see~\ref{sec:PNMOS}), but focused on the abundances to which the
RGS data are most sensitive.
Their temperatures and absorption column depths were fixed to the
values derived from the fits to the pn and MOS spectra, which cover a wider
spectral range.
The normalizations (i.e., emission measures) were left as free parameters to
compensate for potential uncertainties in the cross-calibration.

To fit the RGS data, we grouped them into bins comprising 10~channels, resulting
in an equally-spaced wavelength axis at the cost of
leaving the $\chi^2$-statistic inapplicable. Therefore, we reverted to XSPEC's
W-statistic -- applied when \texttt{c-stat} is used with
Poisson-distributed background --,
which does not rely on approximately Gaussian count distributions in the bins.
Figure~\ref{fig:RGS1_301} shows the merged RGS~$1+2$ spectrum of
\hd\ observed on May 19, grouped at 15~counts per bin for better
visibility. The most prominent spectral feature is the
Lyman-$\alpha$ line of O~\ion{VIII}\ at about
$19$~\AA; further lines of \ion{Ne}{X}, \ion{O}{VIII}, and likely
\ion{Fe}{XVII} are clearly distinguished (see Fig.~\ref{fig:RGS1_301}). The
remaining structure is mostly due to emission of highly ionized iron
ions -- mainly \ion{Fe}{XVIII} and \ion{Fe}{XIX}. The data from RGS~1 and 2 as
well as both spectral orders were fitted simultaneously, and the results are
shown in Table~\ref{tab:PNMOSfit}.

While our fits to the RGS data yielded
higher abundances than those obtained from the MOS and pn cameras, the trend for
high-FIP elements to show higher abundance values is well reproduced. We note
that the abundances of C, N, and O may be modified in the giant due to
dredge-up of nucleosynthesis products from the
stellar interior. Following the first dredge-up, for instance, a rise in the
nitrogen surface abundance accompanied by a decrease in the carbon abundance is
expected \citep{Iben1983}.
There is, however, no clear sign of such a pattern
in our data.

\subsection{ROSAT data and long-term variability}
\label{sec:ROSAT}
An X-ray source with a count rate of $0.294\pm0.023$~\cts\ at the position of
\hd\ has also been detected by ROSAT. Assuming an interstellar
absorption column of $3\times 10^{20}$~cm$^{-2}$ and a mean coronal temperature
between $1$ and $1.5$~keV, we converted the ROSAT count-rate into
unabsorbed $0.3-9$~keV fluxes, $\log_{10}(\mbox{Flux [cgs]})$, between
$-11.5$ and $-11.36$, corresponding to X-ray luminosities between $2.3$
and $4\times 10^{31}$~\ergs.
These estimates are compatible with the flux measurements by \xmm\ and suggest
that the X-ray luminosity of the \hd\ system does not undergo strong variations
on the timescale of decades.

\section{Flares in the \kep\ light-curve}

The \kep\ telescope observed \hd\ uninterruptedly for almost four years
from Quarter~1 through 16 \citep[e.g.,][]{Koch2010}.
While only long-cadence data with about $30$~min sampling have been taken during
the first six quarters, short cadence data, sampled at a temporal cadence of about one
minute, become available for subsequent quarters.
The \kep\ light-curve of \hd\ shows a wealth of intriguing features including
marked transits caused by mutual eclipses of the giant and the dwarf-binary,
transits in the dwarf-binary system, oscillations on the giant, ellipsoidal
variations, rotational modulation, and flares;
a detailed analysis of many of these features can be found in
\citet{Derekas2011} and \citet{Borkovits2013}.

\begin{figure}[h]
  \centering
  \includegraphics[width=0.4\textwidth]{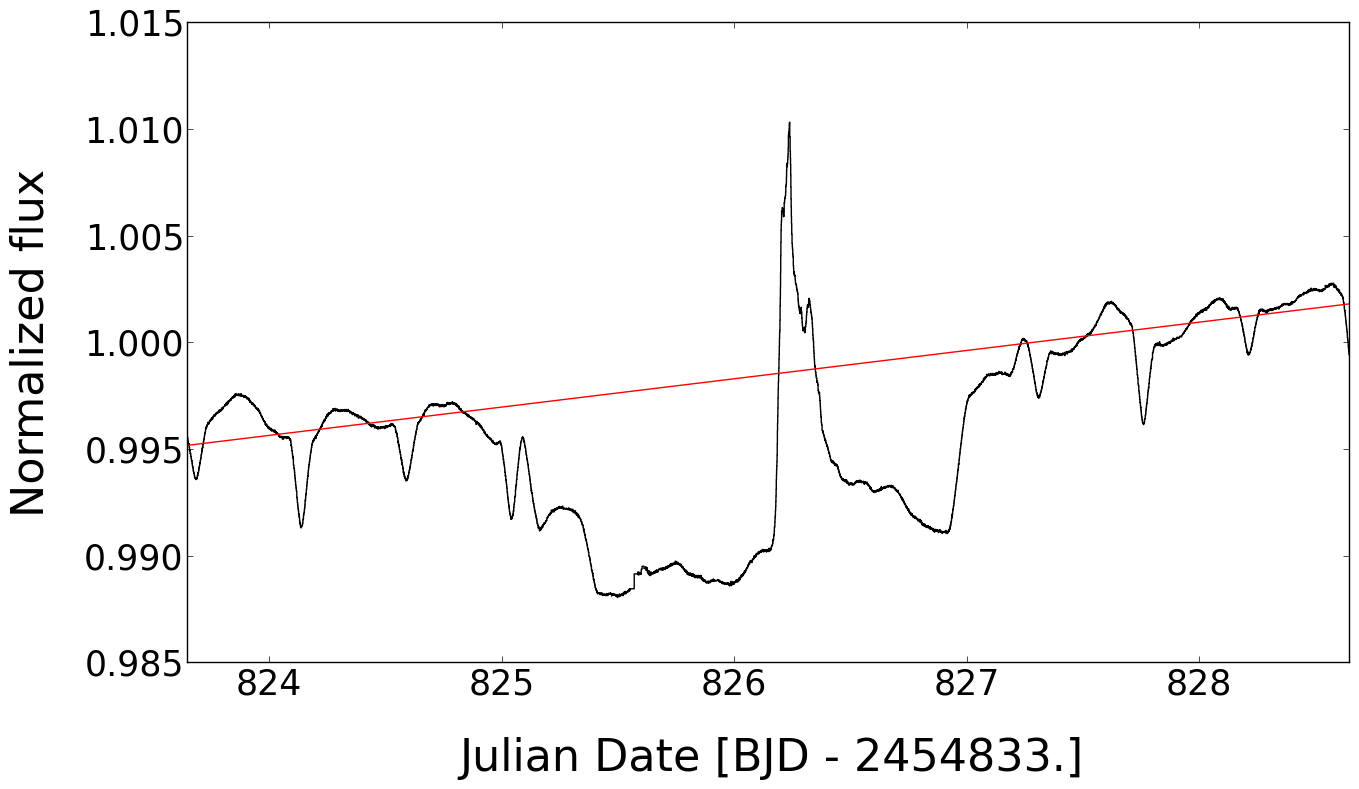} \\
  \includegraphics[width=0.4\textwidth]{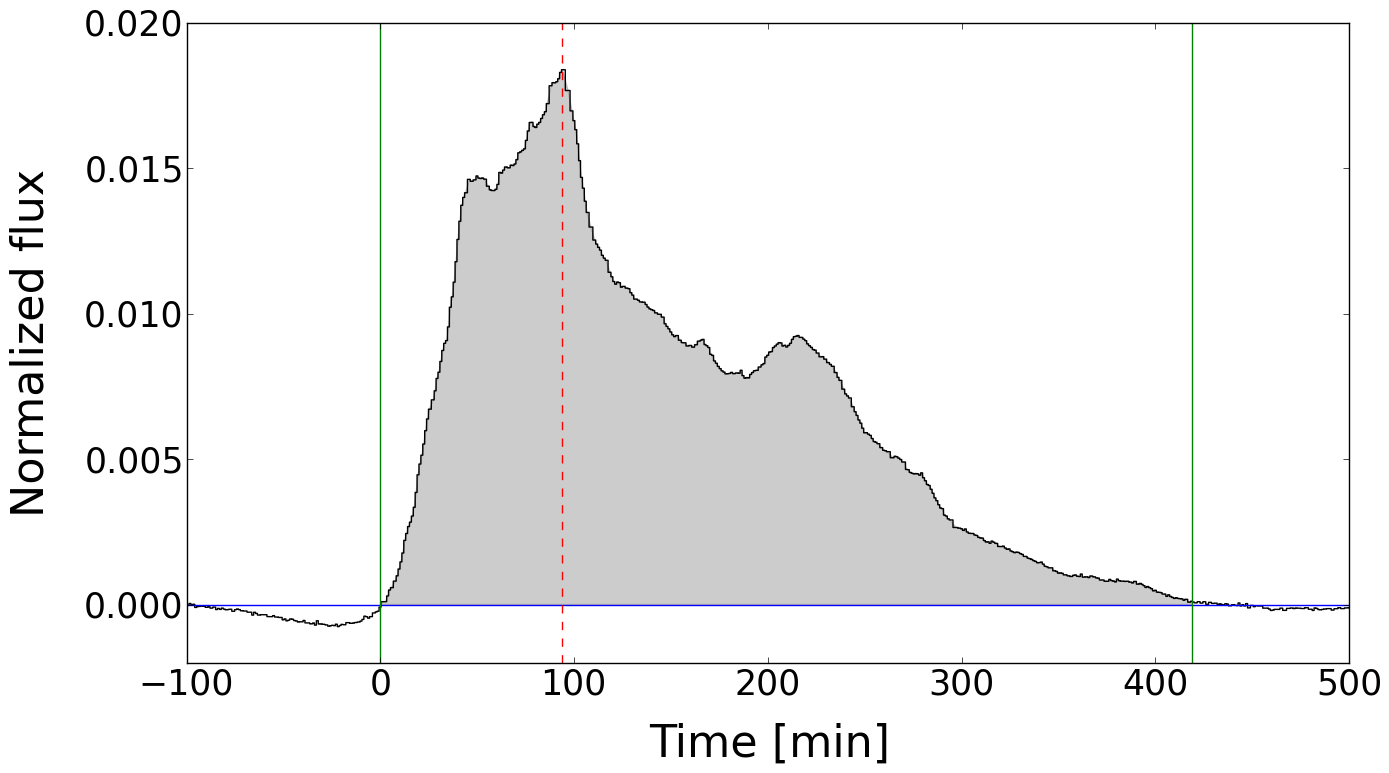}
  \caption{White-light flare observed during secondary eclipse (first flare
           listed in Table~\ref{tab:flares}).
           \textit{Upper panel}: \textit{Kepler} light-curve
           of the secondary eclipse covering the flare.
           The red line shows the continuum normalization of the transit.
           \textit{Lower panel}: Normalized flare light-curve.
           Vertical solid (green) lines indicate the start and end of the flare
           and the dashed (red) line marks the time of maximum flux.
           \label{fig:Kepler-flare-large}}
\end{figure}

\begin{figure*}[t]
  \centering
  \includegraphics[width=0.32\textwidth]{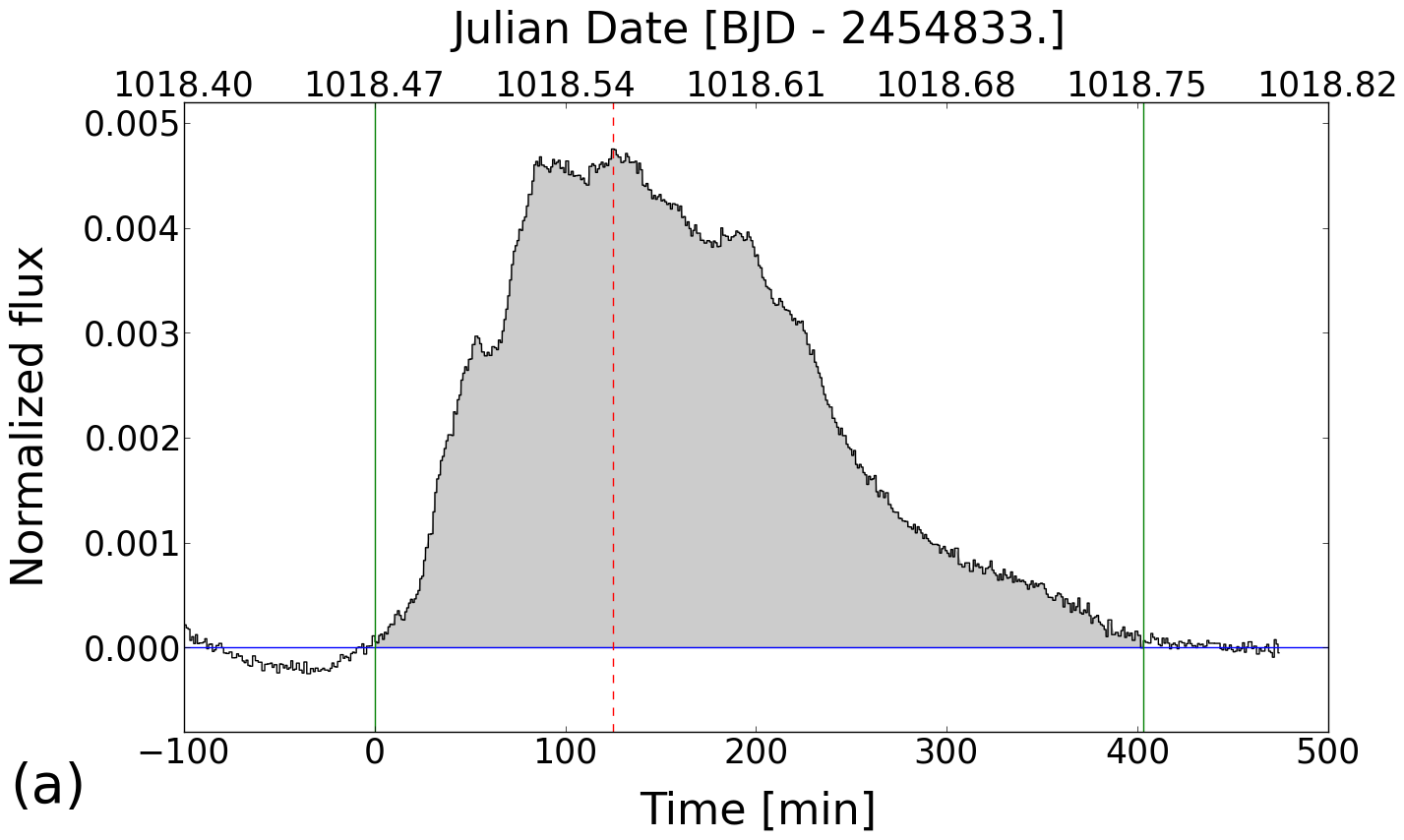}
  \includegraphics[width=0.32\textwidth]{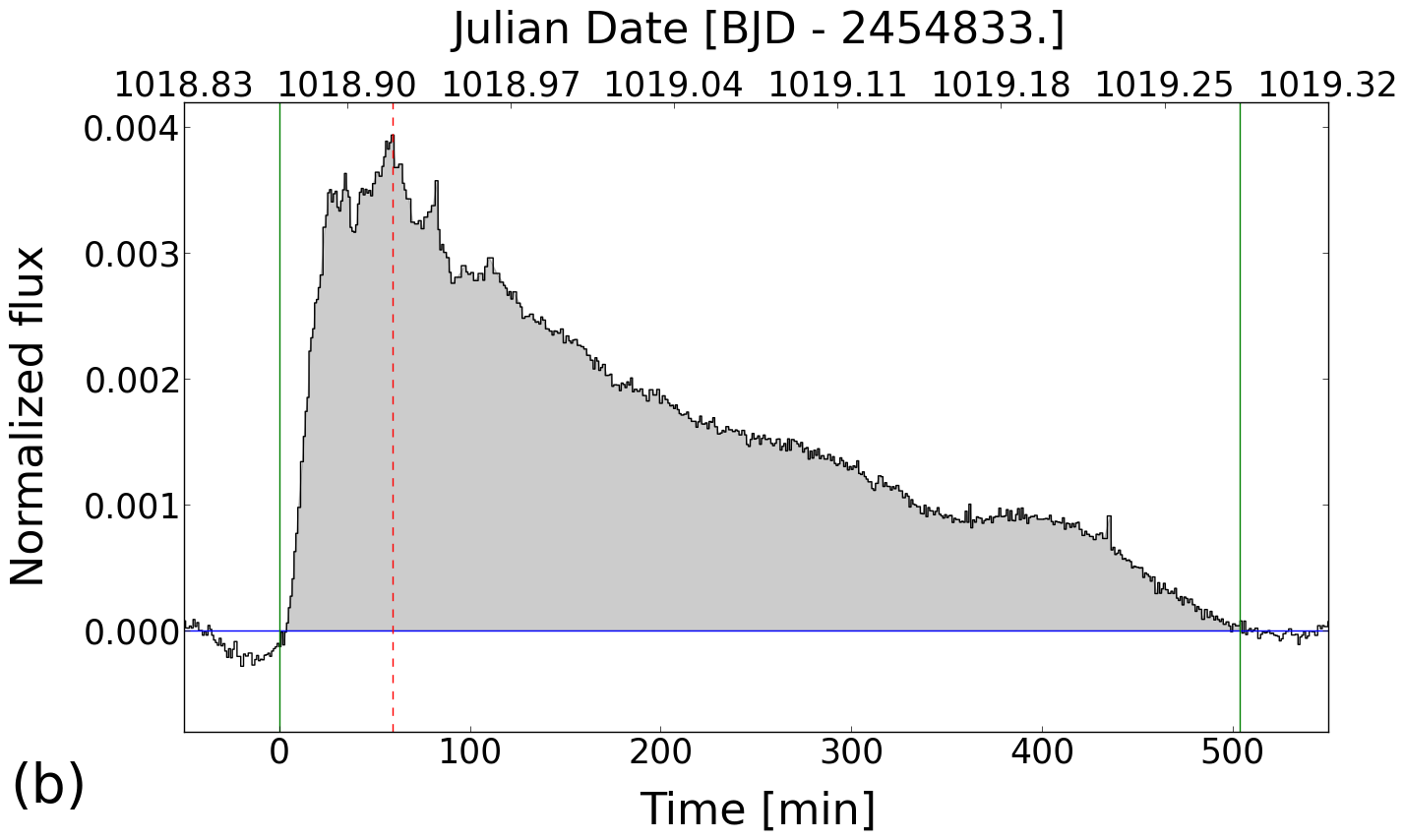}
  \includegraphics[width=0.32\textwidth]{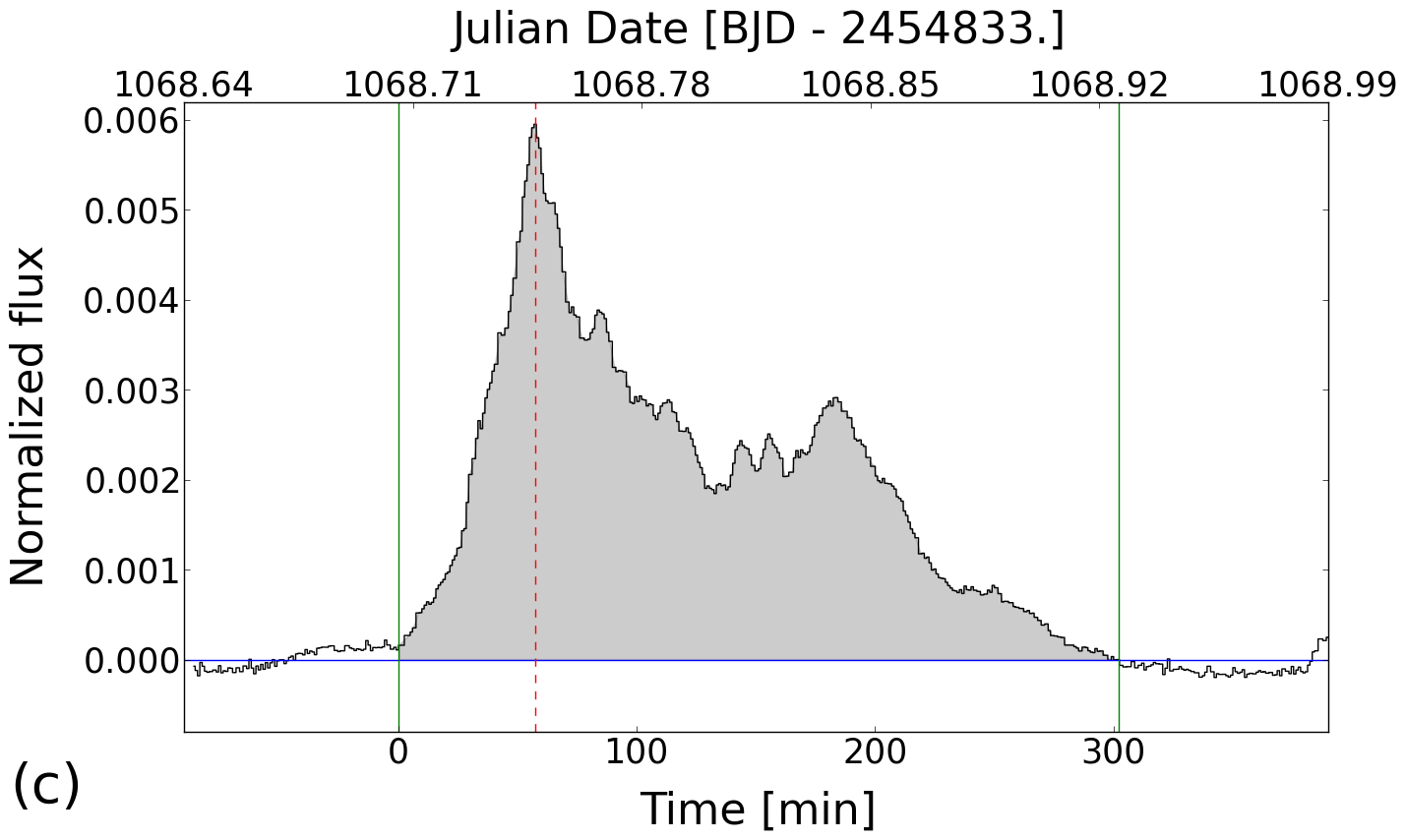} \\[1mm]
  \includegraphics[width=0.32\textwidth]{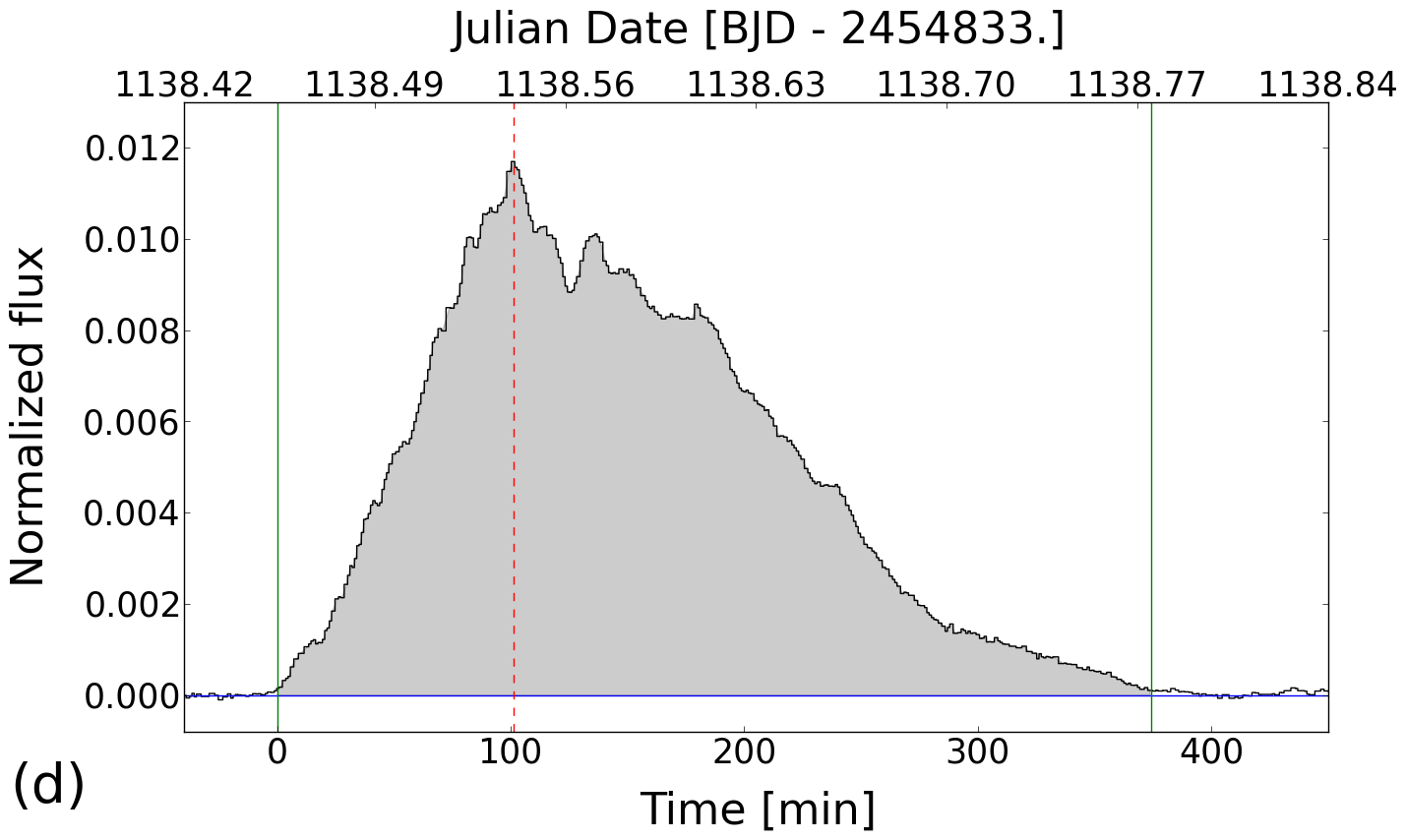}
  \includegraphics[width=0.32\textwidth]{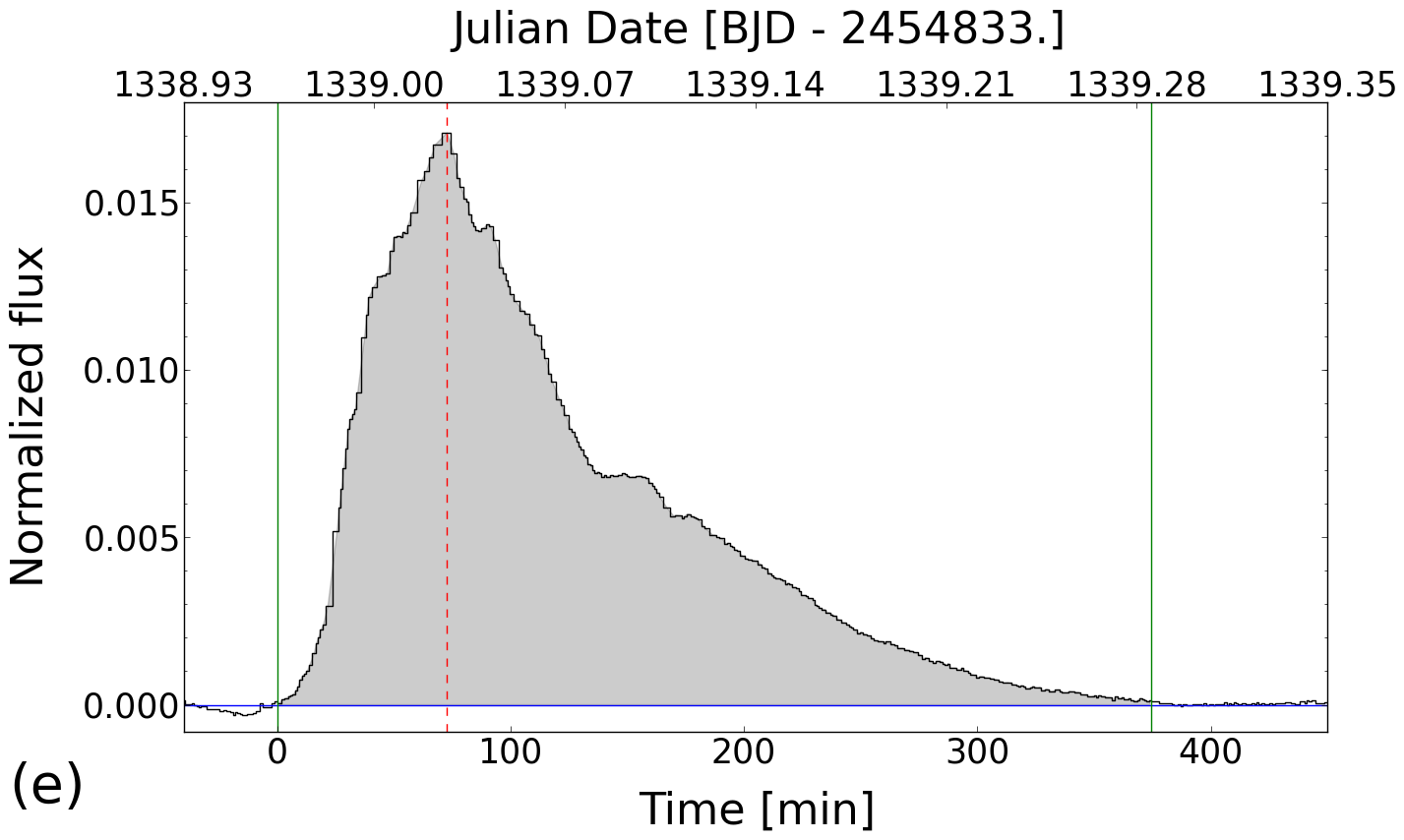}
  \includegraphics[width=0.32\textwidth]{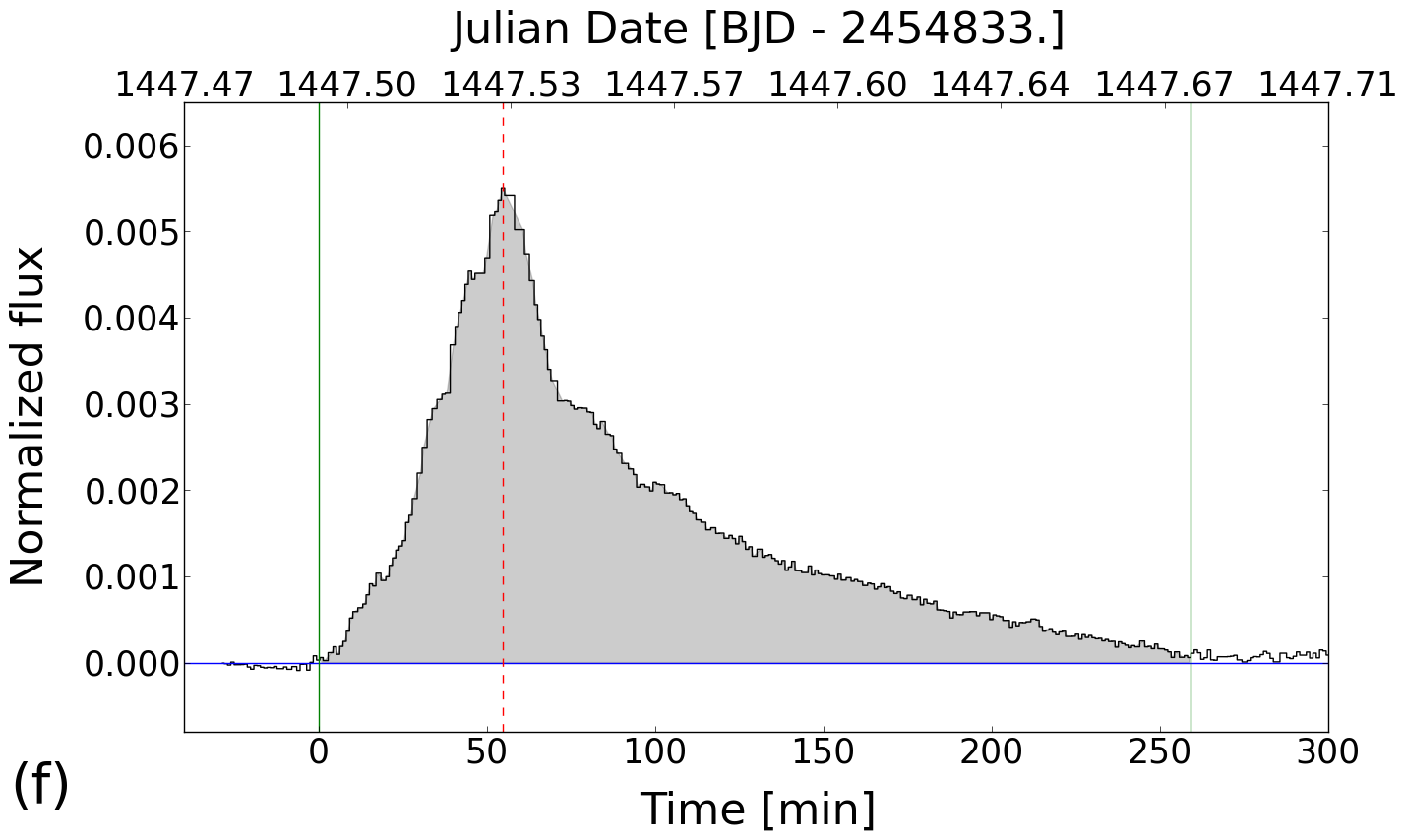}
  \caption{White-light flares observed by \kep. The solid (green) lines indicate
  the start and end of the flares and the dashed (red) line the point of maximum
  flux.
           \label{fig:Kepler-flares}}
\end{figure*}

In our analysis, we concentrated on the short-cadence data, which we
obtained from the \kep\ archive.
We decided to use the Simple Aperture Photometry (SAP) and removed all
data points flagged by the \texttt{SAP\_QUALITY} column.
To search for flares, we visually inspected the resulting light-curve, and thus
identified seven clear flares, whose continuum-normalized light-curves are shown
in Figs.~\ref{fig:Kepler-flare-large} and \ref{fig:Kepler-flares}. We
note the apparent dip in brightness preceding some flares is not physical but
rather caused by the normalization and the rather curved continuum. Although
there are several smaller features in the light-curve that may also be flares,
we found it impossible to verify their true nature, because the system's
optical light-curve is intrinsically strongly variable, and there may be
additional instrumental effects.
Therefore, we restrained our analysis to the seven evident flare events.

Although \citet{Derekas2011} and \citet{Borkovits2013} did not explicitly
investigate the flares, one particularly interesting example has already been
depicted by \citeauthor{Borkovits2013}. This flare -- shown in
Fig.~\ref{fig:Kepler-flare-large} --
occurs during the secondary eclipse in the AB system,
i.e., with the dwarf-binary hidden behind the giant. It can, therefore,
uniquely be ascribed to the primary giant component.

To estimate the flare energy, we applied a continuum normalization to the
individual flares. In particular,
we defined the start and end point of the flares by visual inspection
and fitted a first order polynomial to the flanking parts of the light curve.
The flare light-curves shown in Fig.~\ref{fig:Kepler-flares}
were normalized by the thus obtained polynomial before subtracting unity.
In the case of the flare occurring during secondary eclipse (see
Fig.~\ref{fig:Kepler-flare-large}), we also modeled the transit shape
based on the average secondary-transit profile of the
AB system \citep{Mandel2002}. After dividing by the mean profile, we
performed the same continuum normalization as for the other flares.

In Table~\ref{tab:flares},
we give the time of the flare maximum, T$_\mathrm{peak}$,
the fractional increase in flux at flare maximum,
f$_\mathrm{peak}$, the duration of the rise- and decay-phase,
t$_\mathrm{rise}$ and t$_\mathrm{decay}$, and the total flare energy released in
the \kep\ band.
We defined the duration of the rise phase as the estimated time of its start to
the time of flux maximum. Equivalently, the decay phase lasts from the point of
flux maximum to the estimated end of the flare. Because it is difficult
to clearly distinguish flare-induced variability in the light curve, we
estimated an uncertainty of about $10$~min for these numbers. The released
energy, E$_\mathrm{flare}$, was measured by integrating the flare light-curve and
multiplying the result by the giant's bolometric luminosity (see
Table~\ref{tab:starProp}). Here, we implicitly assumed that
the stellar and the flare spectrum are identical.
Taking into account \kep's sensitivity curve \citep{Cleve2009} and assuming
black-body spectra with a temperature of $10\,000$~K for the flare and $5100$~K
for the star, we found that the estimates would increase by about $30$\%, which
gives an idea of the systematic uncertainties involved in the calculation.

\begin{table}
  \caption{Properties of the seven flares found in the \textit{Kepler}
  short-cadence light-curve of HD~181068. 
  \label{tab:flares}}
  \begin{tabular}{l l l l l}
  \hline\hline
  T$^a_\mathrm{peak}$ & f$^b_\mathrm{peak}$ &
  t$^c_\mathrm{rise}$ & t$^c_\mathrm{decay}$ & E$^d_\mathrm{flare}$ \\
  $[$d$]$ & [\%] & [min] & [min] & [erg] \\ \hline
  $826.238$  & $1.84$ & $94$  & $364$ & $6\cdot10^{38}$ \\
  $1018.557$ & $0.48$ & $125$ & $278$ & $2\cdot10^{38}$ \\
  $1018.941$ & $0.39$ & $59$  & $445$ & $2\cdot10^{38}$ \\
  $1068.750$ & $0.60$ & $57$  & $245$ & $1\cdot10^{38}$ \\
  $1138.560$ & $1.17$ & $101$ & $273$ & $4\cdot10^{38}$ \\
  $1339.051$ & $1.71$ & $73$  & $302$ & $4\cdot10^{38}$ \\
  $1447.538$ & $0.55$ & $55$  & $204$ & $8\cdot10^{37}$ \\
  \hline
  \end{tabular}
  \tablefoot{$^a$\; Time of flare peak in BJD$-$2454833.
             $^b$\; Flare maximum in fraction of stellar flux.
             $^c$\; Duration of rise (between beginning and maximum)
             and decay (between maximum and end) phase of the flare.
             \mbox{$^d$\; Energy} of the flare in white light (\textit{Kepler}
             bandpass).}
\end{table}

\subsection{Third-light origin of the flares?}
Along with the reduced light-curves,
the Kepler data archive provides pixel-scale data with a spatial
resolution of about $4$~arcsec per pixel.
Pixel-center shifts can be used to identify brightness variations not associated
with the source under consideration, but originating on some unrelated back- or
foreground object \citep{Batalha2010}.

To check the source position,
the Kepler pipeline provides ``moment-derived column and row centroids''
(\texttt{MOM\_CENTR1/2}) along with
a correction based on reference stars (\texttt{POS\_CORR1/2}).
In the case of \hd, we found pixel-shifts, which are
essentially proportional to the flux. With about $7$ Kepler magnitudes,
\hd\ is a particularly bright source suffering from charge bleeding and,
therefore, has been assigned a large aperture elongated in the direction of
the bleeding. Our analysis showed that this is also the direction in which the
pixel-shifts are most pronounced, and we conclude that the observed shifts are
most likely due to the characteristics of charge bleeding. In
particular, we found no evidence for the flares being due to a third light
source.

\section{\ion{Ca}{II}~H and K observations of \hd}
\label{sec:CaII}

On Oct. 3, 2013, we carried out a $30$~min observation of \hd\ using the
``Telescopio Internacional de Guanajuato, Rob\'otico-Espectrosc\'opico''
(TIGRE) -- a $1.2$~m telescope located at La Luz, Mexico ($21^{\circ}$~N,
$259^{\circ}$~E) \citep{Schmitt2014}. The telescope is equipped with the
``Heidelberg Extended Range Optical Spectrograph'' (HEROS), a fiber-fed, two-armed instrument, which
provides spectral coverage from $3800-8800$~\AA\ at a resolution of about
$20\,000$.

\begin{figure}[h]
  \includegraphics[angle=-90, width=0.49\textwidth]{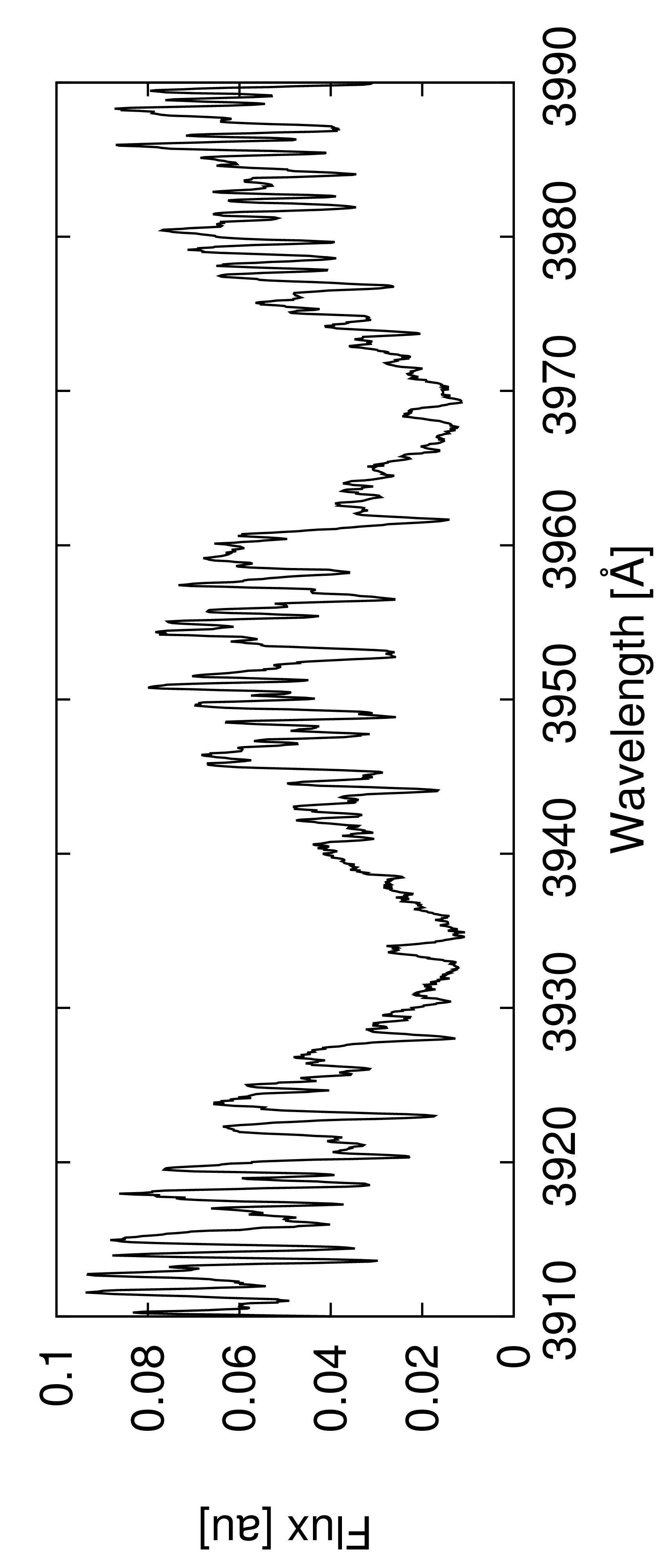}
  \caption{Ca~\ion{II}~H and K lines of \hd~A observed by TIGRE.
  \label{fig:TigreCaHK}}
\end{figure}

The exposure was scheduled during secondary
eclipse, so that only the giant's spectrum has been observed.
Figure~\ref{fig:TigreCaHK} shows the \ion{Ca}{ii}~H and K lines with
pronounced chromospheric fill-in, which has already been noticed by
\citet{Derekas2011}, who used the width of the emission cores to estimate the
stellar absolute brightness via the Wilson-Bappu effect. From our spectrum, we derived 
a Mount-Wilson S-index of $0.41\pm0.02$.

Using the relations
given by \citet{Noyes1984} and the color-dependent conversion factor
($C_{cf}$) given by \citet{Rutten1984}, we converted the S-index into
a $\log(R'_{HK})$ value of $-4.57 \pm 0.01$.
Note that the conversion between Mount-Wilson S-index and
$\log(R'_{HK})$ has been revisited by a number of authors such as \citet{Hall2007}
and \citet{Mittag2013}, who obtained results differing by up to a
factor of about two; see \citet{Hall2007} for a discussion of the development.
For instance, using the calibration of the ``arbitrary
units'' provided by \citet{Hall2007}, we arrive at $\log(R'_{HK})$ ratio of $-4.41$,
which gives an impression of the systematic uncertainty involved in the
conversion.

\begin{figure}[h]
  \includegraphics[angle=-90, width=0.49\textwidth]{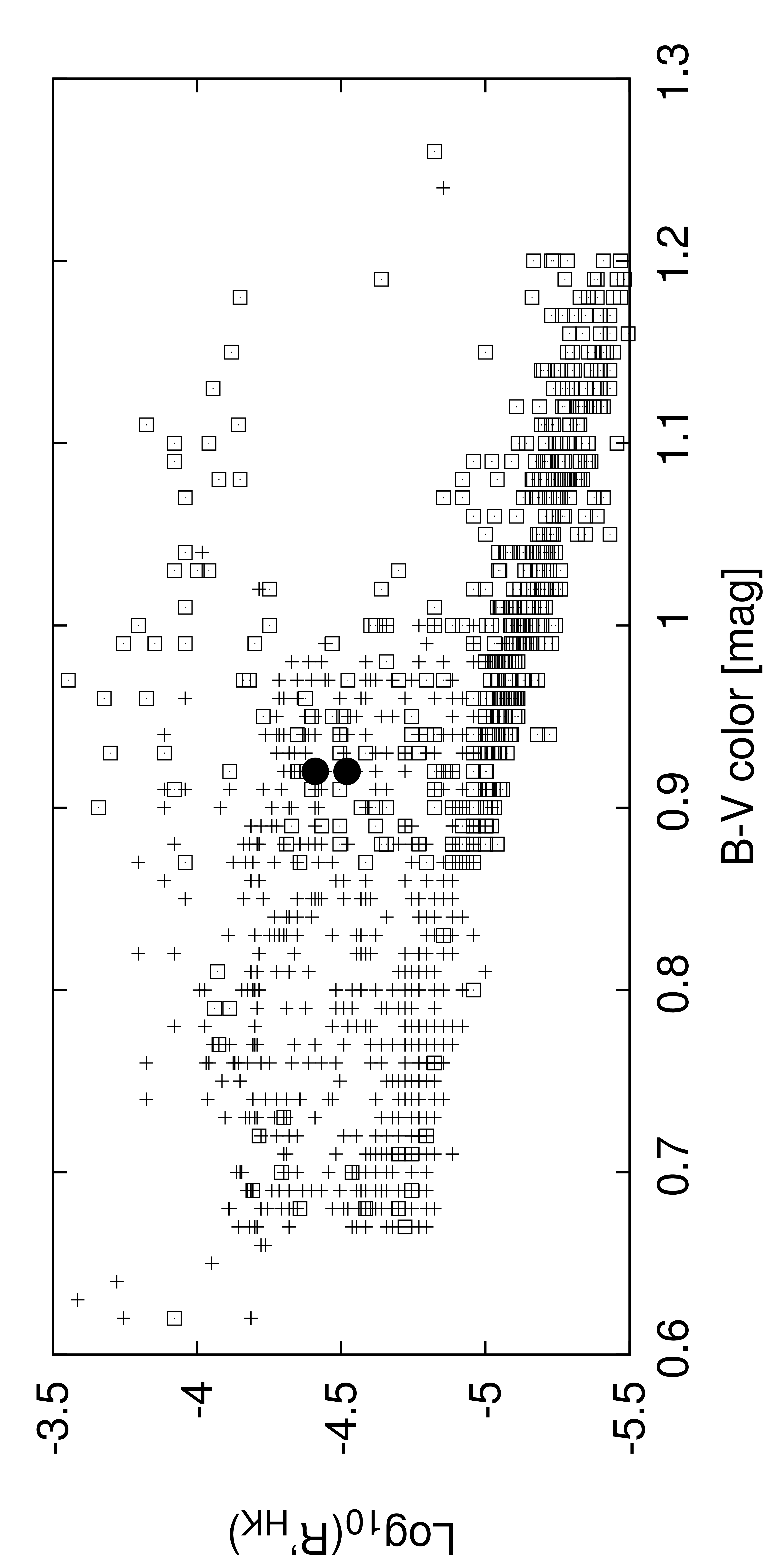}
  \caption{The $\log(R'_{HK})$ index of \hd\ (filled circles) with the
  calibration of \citet{Noyes1984} and \citet{Hall2007} in comparison to the indices
  measured in the sample presented by \citet{Strassmeier2000}; pluses denote
  main-sequence stars and squares subgiants and giants.
  \label{fig:LogRHK}}
\end{figure}

Its chromospheric emission puts \hd~A among the
active giants with a chromosphere clearly emitting \ion{Ca}{ii}~H and K
emission-line cores in excess of the basal level, which corresponds to an S-index value of
$\lesssim 0.15$ for a G-type giant \citep[][Fig.~3]{Duncan1991, Schroeder2012}.
Figure~\ref{fig:LogRHK} shows the $\log(R'_{HK})$ index of \hd\ in comparison to
measurements of main-sequence and (sub-)giant stars presented by
\citet{Strassmeier2000}. Independent of the details of the calibration, \hd\
qualifies as an active, albeit not an outstandingly active, giant star.
We caution, however, that a single observation provides only a snapshot
of the chromospheric properties.

\section{Discussion}

\subsection{Origin of the X-ray emission}

Our detection of strong X-ray emission from \hd\ during both the in- and
out-of-eclipse phase suggests that the bulk of X-ray emission originates on the
giant primary component. With the observed X-ray luminosity being even higher
when the dwarf-binary was eclipsed, it remained impossible to directly identify
the contribution of the dwarfs to the overall X-ray flux.

To estimate the dwarf-binary's contribution to the total X-ray luminosity, we
assumed coronal emission at the saturation limit of $L_X/L_{bol}=10^{-3}$,
which has also been observed in other fast-rotating low-mass binaries such as
YY~Gem \citep{Tsikoudi2000,
Stelzer2002, Pizzolato2003}.
Considering a total bolometric
luminosity of $0.72$~L$_{\odot}$ for the dwarf-binary in \hd, we estimated an
upper limit of $2.8\times 10^{30}$~\ergs\ for their X-ray luminosity.
Consequently, the dwarf-binary could contribute up
to $15$\% of the quiescent X-ray flux of the \hd\ system. Because this
contribution cannot be specified any further and the
giant provides at least $85$\% of the X-ray flux, we attribute the entire X-ray
emission to the giant during the remaining analysis.

\subsection{\hd\ as an \rs\ system}

\rs\ systems are among the strongest coronal X-ray sources in the Galaxy.
Compared to single late-type giants in the solar neighborhood \citep{Huensch1996},
the X-ray luminosity of \hd\ is elevated by about two orders of magnitude.
\citet{Dempsey1993} and \citet{Dempsey1993b} studied the X-ray
emission of \rs\ systems based on ROSAT data.
In their analysis of the coronal temperature, \citet{Dempsey1993b} found a bimodal
distribution with peaks around $2\times 10^{6}$~K and $1.6\times
10^7$~K.
Although we determined higher temperatures in the
quiescent phase of \hd\ (i.e., the May 3 observation), we also found
satisfactory fits based on a two-temperature model in our analysis. This
model certainly remains an approximation
to the true distribution of emission measure in \hd, given that studies of the differential
emission measure (DEM) in other \rs\ systems based on grating spectroscopy have
revealed a continuous distribution of the emitting plasma
\citep[see ][]{Huenemoerder2001, Drake2001, Huenemoerder2003}. However, the
reconstructed DEMs also show distinct peaks \citep[][Fig.~5]{Huenemoerder2003},
so that a two-component approximation could, indeed, coarsely characterize the
DEM. Attributing the observed X-ray emission to the giant, we estimated a
quiescent surface X-ray flux, $F_X$, of $2\times 10^{6}$~\ergcms.
The resulting ratio of $0.57$ of \ion{Ca}{II}~H and K line surface-flux
(Sect.~\ref{sec:CaII}) to $F_X$ is compatible with the values observed in the
sample of young solar analogs presented by \citet{Gaidos2000} or the sample of
giants shown by \citet{Pallavicini1982}.

With its quiescent X-ray
properties, we found that \hd\ fits nicely into the sample presented by
\citet[][Figs.~3 and 4]{Dempsey1993}. 
Assuming synchronous rotation of the giant and the AB system, we found
that \hd\ does neither show an outstanding X-ray luminosity nor surface X-ray flux.
Also the relation between stellar radius and rotation period observed in \hd\
does not set it apart from the sources reported by \citet{Dempsey1993}.
Finally, 
we used the method presented by \citet{Eggleton1983} to compute a value of
$0.4$ for the giant's Roche-lobe filling factor, $R_s/R_{lobe}$, and compared it
with its X-ray luminosity and surface X-ray flux; in both cases, the result is
compatible with the sample properties presented by
\citet{Dempsey1993} (see their Fig.~5).

\citet{Dempsey1993}
concluded that ``other than acting to tidally spin up the system, the secondary
plays no direct role in
determining the X-ray activity level.'' While the oscillation spectrum of \hd~A
does clearly show that there is an interaction between the orbiting
dwarf-binary and the giant's atmosphere, this conclusion is not otherwise
challenged by our X-ray observations of \hd.

\subsection{Coronal abundances and first-ionization-potential effect}
\citet{Derekas2011} determined a subsolar photospheric metallicity for \hd.
Depending on the applied method, the authors state values of $[M/H] =
-0.6\pm0.3$ and $[M/H] = -0.2\pm0.1$. These numbers are compatible with the mean
coronal abundance for the low- and medium-FIP elements, which amounts to about
$0.3$ times its solar equivalent corresponding to $[X/H] \approx -0.5$ for the
respective element $X$ (Table~\ref{tab:PNMOSfit}).

The coronal abundance of the high-FIP elements is approximately solar, making
them about three times overabundant with respect to the low- and medium-FIP
elements. Assuming an overall solar photospheric abundance pattern in \hd, the
abundance distribution is characteristic of an ``inverse first
ionization-potential'' (IFIP) effect. While the IFIP effect is typically
observed in the coronae of active stars \citep[e.g.,][]{Brinkman2001, Drake2001,
Telleschi2005}, inactive stars tend to show a solar-like FIP-effect. The coronal abundances
of \hd, therefore, indicate an active star.

\subsection{White-light and X-ray flares}

We analyzed seven flares observed by
\kep. Some of the flares show profiles indicative of multiple flare
events (e.g., Figs~\ref{fig:Kepler-flare-large} and \ref{fig:Kepler-flares}~c),
which may result from the eruption of a cascade of magnetic loops. In the
following, we discuss their origin in the \hd\ system, their energetics, and
potential impact on coronal heating.

\subsubsection{Location of the flares in the \hd\ system}

Because our analysis yielded no evidence for a third light responsible for the
flares, we attribute them to the \hd\ system, where they may, in
principle, be located on both the active giant and the dwarf-binary system.
\citet{Shibayama2013} have analyzed 1547 superflares on G-type dwarfs identified
in the \kep\ data. The strongest flare detected by these authors released an
energy of $1.3\times 10^{36}$~erg, which remains about two orders of magnitude
below the flares observed on \hd. This suggests that all
flares are associated with the giant rather than the dwarf-binary.

The peak luminosities of the flares recorded on \hd\ in the \kep\ band
reach between $0.39$\% and $1.8$\% of the giant's luminosity,
which corresponds to $0.36$ to $1.7$~L$_{\odot}$ assuming the same
spectrum (Tables~\ref{tab:starProp} and \ref{tab:flares}).
Therefore, the peak flare-luminosity is comparable to or even higher than the
total luminosity of
the dwarf-binary ($0.7$~L$_{\odot}$). Assuming that the flaring material
seen by \kep\ had a temperature of $T_{eff,f} = 10\,000$~K
\citep[][]{Neidig1989, deJager1989, Hawley2003} and the dwarf stars have radii,
$R_s$, of $\approx 0.8$~R$_{\odot}$, we used the expression
\begin{equation}
  f = 2 \frac{L_{peak}}{\sigma T_{eff,f}^4 4\pi R_s^2}
\end{equation}
to estimate
that between $12$\% and $55$\% of the
visible hemisphere of either dwarf star would have had to be covered by flaring
material to reproduce the flare peaks observed by \kep. Such filling factors appear
unreasonably large compared, e.g., to filling factors of $\approx 3$\%
derived for an unusually intense flare observed on EV~Lac \citep{Osten2010}. In
contrast, the filling factor required on the giant amounts to only about
$0.05-0.2$\%. This again favors the giant as the origin of the flares, which
produces also the bulk of the observed X-ray emission.

It has been speculated that particularly strong flare events in binary systems
originate in an interbinary filament connecting the binary components.  
Such an interbinary scenario has, e.g., been
proposed to be responsible for an enormous X-ray outburst in the \rs\ system HR~5110
\citep{Graffagnino1995}. However, \citet{Schmitt1999} could confine the location
of another giant X-ray flare in the Algol system to the B-component, ruling out
an interbinary location of the X-ray emitting material. 
One of the flares observed in \hd\ can uniquely be ascribed to the giant,
because it was observed during secondary eclipse, when the dwarf-binary
remained invisible (see
Fig.~\ref{fig:Kepler-flare-large}). This renders an interbinary origin
of the flare
unlikely, because both the interbinary space and the giant's
photosphere facing the dwarf-binary remain hidden. Therefore, we argue
that the flaring material should rather be confined to the surface of the
giant.

In their analysis of the frequency spectrum of the 
\kep\ light-curve, \citet{Borkovits2013} detected three related peaks of which
the first is located at a frequency of $2.208\,29$~d$^{-1}$, i.e., exactly half
the eclipse period of the dwarf-binary; the frequencies of the other two are a
combination of the periods of the dwarf-binary and the AB system. With
such a configuration -- the authors state -- a tidal origin of the associated
small-amplitude oscillation on the stellar surface is ``out of question''.
Therefore, we searched for a relation
between the flare timing and both the period of the AB- and dwarf-binary
systems.
Given our current set of seven flares, we could not find any tangible relation
between the orbit phase of the AB- or dwarf-binary system and the flares,
however.
This is compatible with flares erupting on the giant unrelated to the
orbital configuration of the system.

\subsubsection{The elevated level of X-ray emission on May 19}

\citet{Pandey2012} studied X-ray emission and flares on five \rs\ systems using
\xmm\ data. For these systems, the authors derived quiescent X-ray luminosities
between $5\times 10^{30}$~\ergs\ and $8\times 10^{30}$~\ergs. Their sample
comprises the highly active \rs\ system UZ~Lib, which consists of a
K0III-type giant and a low-mass companion in a close orbit with a period of $4.76$~d 
\citep{Olah2002}. As reported on by \citet{Pandey2012}, \xmm\ observed UZ-Lib
twice: first, UZ~Lib was caught in a likely quiescent state, where it showed
hardly any variation in its X-ray luminosity for about $15$~ks; second, the
system was observed in a more active state characterized by an approximately
doubled count-rate, decaying, however, at a rate of about
$-0.03$~ct\,s$^{-1}$\,ks$^{-1}$ for at least $28$~ks, i.e.,
the duration of the observation. This behavior is
similar to what we observed in \hd\ on May 19 and likely the consequence of
having caught the system in the decay phase
of a flare with unobserved rise- and peak-phase \citep{Pandey2012}.
This interpretation is also compatible with the decrease in hardness
seen during the observation.

Had the X-ray count-rate of \hd\ continued to decline with the
rate observed on May 19, the count-rate level observed on May 3 would have
been reached after about $150$~ks or $1.8$~d. In this period,
$0.5\times (5.9-1.9)\times 10^{31}\mbox{~\ergs} \times
150\mbox{~ks}$ $=3\times 10^{36}$~erg of energy would have been released
in excess of the quiescent X-ray emission, which
corresponds to $\approx 1$\%
of the typical energy released in the strong white-light flares. The total
amount of energy released in the hypothesized X-ray flare was probably
substantially higher, but can hardly be estimated as the most intense phase
remained unobserved. For comparison, the strongest flare reported on by
\citet{Pandey2012} released about $4.2\times 10^{37}$~erg in X-rays.

If only the decay phase of a potential soft X-ray
flare on \hd\ was observed, the duration of the entire flare must have exceeded
the exposure time of $15$~ks (i.e., $250$~min).
This may be compared to the typical duration
of $400$~min for the white-light flares, which are probably associated with the
impulsive flare-phase and, thus, are believed to precede the longer-lasting soft X-ray
flare \citep[e.g.,][]{Neidig1993}.
Long-duration X-ray flares have also been observed in other \rs\
systems. To our knowledge, the current record holder is a 9~day flare event
observed by ROSAT on CF~Tuc \citep{Kuerster1996}. Attributing the elevated
count-rate level of \hd\ to a flare is, therefore, plausible from both the
point of view of energy budget and timing. Nonetheless, alternative
explanations such as an elevated quasi-quiescent level cannot be ruled out.

\subsubsection{Coronal heating by flares}
To determine whether coronal flare heating could potentially account for
the observed X-ray emission, we studied the flare energetics.
The total amount of energy released in the observed white-light flares
amounts to $17.8\times 10^{38}$~erg (see Table~~\ref{tab:flares}) with
individual flare energies ranging from $0.8$ to $6\times10^{38}$~erg. As the
analyzed short-cadence data cover about $928$~d, this translates into a mean
rate of $2.2\times 10^{31}$~\ergs\ of flare energy released in white-light. This
value remains a lower limit, because, most likely, we observed only the upper end
of the flare energy distribution.

X-ray and EUV studies have shown that the distribution of the flare
rate, $N$, as a function of flare energy, $E$, obeys a power-law of the form
\begin{equation}
  \frac{\partial N}{\partial E} = k_1 E^{-\alpha} \; ,
\end{equation}
where $k_1$ is a constant \citep[e.g.,][]{Collura1988, Audard1999, Osten1999,
Audard2000}. \citet{Shibayama2013} show that the same holds for white-light
superflares. Assuming \mbox{$\alpha = 2$}, as suggested by
\citet{Shibayama2013}, we integrated from the lowest to the highest flare
energy reported in Table~\ref{tab:flares} to determine the constant, $k_1$, to
be \mbox{$8.8\times 10^{38}$~erg\,$(928$~d)$^{-1} = 1.1\times 10^{31}$~\ergs}.
Ultimately, the total energy released in flares cannot be determined
as the unobserved lower energy cut-off is crucial.
Assuming, however, that the flare-energy distribution reaches down to
$10^{30}$~\ergs, white-light flares would release energy at a rate of
$2.2\times 10^{32}$~\ergs. A fraction of $10$\% of that energy radiated
in X-rays suffices to account for the observed quiescent X-ray emission of \hd.
Thus, it seems plausible that coronal
heating by flares could provide a fraction of the observed quiescent X-ray
luminosity.

\subsection{\hd\ as a model system for star-planet-interaction}

\begin{figure}[h]
  \includegraphics[width=0.49\textwidth]{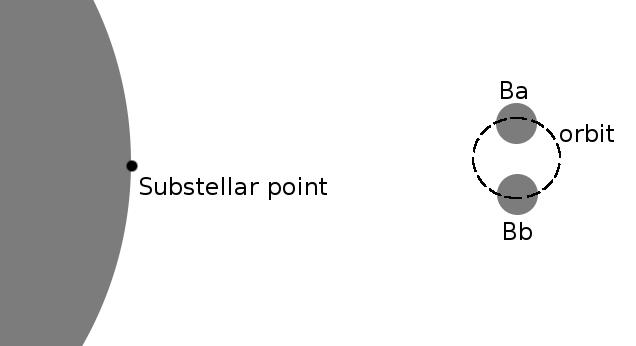}
  \caption{Sketch (not to scale) of the \hd\ system.
  \label{fig:Sketch}}
\end{figure}

The basic geometry of \hd\ is similar to that of a planetary system
with the dwarf-binary representing a double planet or a planet with a giant
moon (see Fig.~\ref{fig:Sketch}). However, the mass ratio of $0.6$ clearly sets
\hd\ apart from usual planetary systems, which show mass ratios on the
order of a few $10^{-3}$.

Assuming that the rotation period of the giant and the orbital period of the
dwarf-binary are synchronized and the rotation axis and orbit normal are
aligned, the position of the dwarf-binary remains fixed in the frame of the
rotating giant. In this configuration, the 
gravitational force exerted by the dwarf-binary on the atmosphere of the giant
can be separated into two components: a stationary component and a
cyclic component. The latter is caused by the (internal) orbital motion
of the revolving dwarf-binary system, which permanently changes its
configuration and, for instance, exerts a different force when seen in
conjunction than in quadrature. In particular, the variable force component at
the substellar point, $\vec{F}_{SP}$, is given by the vectorial sum of the
forces exerted by the dwarf-binary components, $\vec{F}_{Ba,b}$, after
subtracting their time average

\begin{equation}
  \Delta \vec{F}_{SP}(t) = \left(\vec{F}_{Ba}(t) + \vec{F}_{Bb}(t)\right) -
  <\vec{F}_{Ba}(t) + \vec{F}_{Bb}(t)> \; .
\end{equation}

In \hd\ a number of oscillation modes have been detected in the \kep\ light
curve, which are likely driven by this periodic force 
\citep{Derekas2011, Borkovits2013, Fuller2013}.
Using the masses and orbital elements given in \citet{Borkovits2013}, we
calculated that the change in gravitational acceleration on the substellar point
on the giant's surface due to the (internal) orbital motion of the dwarf-binary
amounts to $\approx 3\times 10^{-2}$~cm\,s$^{-2}$ both in radial and tangential
direction.

In the following, we show that the variable force exerted on the surface of the
giant \hd~A is similar to that exerted by a hot Jupiter on the surface of
its host star. Because stellar rotation and orbital motion are typically not
synchronized in planetary systems, the planet moves with respect to the stellar
surface. This produces a variable but nonetheless periodic configuration of the
planetary body and individual stellar surface elements and, thus, a periodic
surface force potentially similar to that seen in \hd.
A particularly intriguing example is a polar orbit during which the
planet passes the stellar poles periodically, independent of the stellar
rotation period \citep[e.g.,][]{vonEssen2014}. This
change in gravitational pull on the stellar atmosphere is among the proposed
mechanisms for star-planet interaction
(SPI), whose reality, however, remains controversial
\citep[e.g.,][]{Shkolnik2008, Scharf2010, Poppenhaeger2010}.

In the HD~189733 system -- a typical planetary system with a hot Jupiter --
the gravitational acceleration caused by the planet HD~189733\,b on the stellar
atmosphere at the substellar point amounts to about $1$~cm\,s$^{-2}$
\citep[see, e.g.,][for the orbital elements]{Bouchy2005}.
Hence, the amplitude in local acceleration on the stellar
surface caused by HD~189733\,b is about a factor of $30$ larger than the
amplitude of the periodic force at the substellar point in \hd. The latter
is, however, sufficient to drive tidally-induced oscillations
\citep{Fuller2013}; extrapolating, a similar interaction may be conceivable in
HD~189733 and other comparable planetary systems.
In contrast to HD~189733, however, \hd\,A is a giant with a highly tenuous
atmosphere and a surface gravity of only $\log(g)=2.73$, i.e., about two orders
of magnitude below that of HD~189733 with $\log(g)=4.53\pm0.14$ \citep{Bouchy2005}.
Yet, in relation to the local gravity, the variable component is
comparable in \hd\ and HD~189733.

We conclude that hot Jovian planets revolving around main-sequence stars
could cause observable changes in the stellar atmosphere analogous to the
tidally driven oscillations observed in \hd. Whether this, however, produces
any change, e.g., in the level or behavior of stellar activity remains unclear.
At least in the case of
\hd, we did not find any such relation: neither an elevated level of X-ray emission
was found compared to other \rs\ systems nor a correlation between the flare
timing and the orbital motion of the dwarf-binary could be identified.

\section{Summary and conclusion}
We have studied stellar activity in the hierarchical triple system \hd, using   
X-ray observations, \kep\ light-curves, and an optical spectrum. With an S-index
of $0.41\pm 0.01$ and a resulting $\log(R_{HK})$ value of $-4.52 \pm 0.01$, 
\hd~A qualifies as an active -- albeit not extremely active --
giant member of an \rs\ system. This finding is compatible with the detection of
quiescent X-ray emission at a level of $\approx 2\times 10^{31}$~\ergs\ originating
predominantly in the giant's corona with a temperature distribution
appropriately represented with a two-component thermal model peaked around
$0.8$~keV and $1.9$~keV, and the presence of an inverse FIP effect.
A comparison with the ROSAT survey,
carried out more than a decade prior to our \xmm\ program, did not reveal strong
variability in the quiescent X-ray emission of \hd.

During the \xmm\ observation on May 19, \hd\ showed a three-fold elevated level
of X-ray emission compared to the quiescent state with a clearly declining
gradient and spectral hardness, which may have arisen from having
observed the system during the decay phase of a flare.
This is compatible with strong white-light flares observed in
the \kep\ light-curves, which likely originate on or close to the surface
of the giant. The observed white-light flares release up to $6\times
10^{38}$~erg of energy in the \kep\ band and potentially contribute
significantly to the coronal heating.
Although the observation on May 19 had been scheduled
during secondary eclipse, the observed intrinsic variability
prevented us from disentangling the contributions of the
giant and the dwarf-binary. Based on coronal saturation, we estimated that
at least $85$\% of the observed X-ray emission must be attributed to
the giant. Our findings are compatible with the hypothesis that strong magnetic
activity suppresses solar-like oscillations in the giant.

As a result of its (internal) orbital motion, the
dwarf-binary companion imposes a cyclic tidal distortion on the giant's
atmosphere, which shows up as an oscillation in the \kep\ light-curve.
Our estimates show that the cyclic gravitational force imposed on the substellar
point of the giant's surface is smaller in amplitude than the change in local
surface gravity caused by a typical revolving hot Jupiter. Therefore, \hd\
may serve as a model system to study star-planet-interactions.
Based on the current sample of seven flares, we could not detect
any correlation between the flare properties and
the orbital motion of the AB- or dwarf-binary system.
Further, the binary nature of \hd~B does not seem to have an impact 
on the giant's bulk X-ray emission or activity if compared
to other \rs\ systems; in fact, the X-ray properties of \hd\ appear to be rather
typical when compared to other \rs\ systems --
a designation also adequate for \hd.

\begin{acknowledgements}
KFH acknowledges support by the DFG under grant HU~2177/1-1. PCS acknowledges
support from the DLR under grant \mbox{DLR 50 OR 1307}. This work is based on
observations obtained with XMM-Newton, an ESA science mission with instruments
and contributions directly funded by ESA Member States and NASA. 
\end{acknowledgements}

\bibliographystyle{aa}
\bibliography{hd181068.bib}

\end{document}